\documentclass[%
reprint,
superscriptaddress,
nofootinbib,
amsmath,amssymb,
aps,
pre,
]{revtex4-1}

\usepackage{graphicx}
\usepackage{dcolumn}
\usepackage{bm}
\usepackage{natbib}
\usepackage{graphicx}
\usepackage{comment}
\usepackage{mathtools} 

\usepackage{float}
\usepackage{hyperref} 
\usepackage{xcolor}
\graphicspath{{figures/}}

\newcommand{\dd}{\mathrm{d}}
\newcommand{\e}{\mathrm{e}}

\begin{document}

\title{Bounded Rationality and Animal Spirits: \\
A Fluctuation-Response Approach to Slutsky Matrices 
}

\author{Jérôme Garnier-Brun}
\email{jerome.garnier-brun@polytechnique.edu}
\affiliation{Chair of Econophysics and Complex Systems, \'Ecole polytechnique, 91128 Palaiseau Cedex, France}
\affiliation{LadHyX UMR CNRS 7646, \'Ecole polytechnique, 91128 Palaiseau Cedex, France}

\author{Jean-Philippe Bouchaud}%
\affiliation{Chair of Econophysics and Complex Systems, \'Ecole polytechnique, 91128 Palaiseau Cedex, France}
\affiliation{Capital Fund Management, 23 Rue de l'Universit\'{e}, 75007 Paris, France}
\affiliation{Académie des Sciences, 23 Quai de Conti, 75006 Paris, France\smallskip}

\author{Michael Benzaquen}%
\affiliation{Chair of Econophysics and Complex Systems, \'Ecole polytechnique, 91128 Palaiseau Cedex, France}
\affiliation{LadHyX UMR CNRS 7646, \'Ecole polytechnique, 91128 Palaiseau Cedex, France}
\affiliation{Capital Fund Management, 23 Rue de l'Universit\'{e}, 75007 Paris, France}

\date{\today}

\begin{abstract}
The Slutsky equation, central in consumer choice theory, is derived from the usual hypotheses underlying most standard models in Economics, such as full rationality, homogeneity, and absence of interactions. We present a statistical physics framework that allows us to relax such assumptions. We first derive a general \textit{fluctuation-response} formula that relates the Slutsky matrix to spontaneous fluctuations of consumption rather than to response to changing prices and budget. We then show that, within our hypotheses, the symmetry of the Slutsky matrix remains valid even when agents are only boundedly rational but non-interacting. We then propose a model where agents are influenced by the choice of others, leading to a phase transition beyond which consumption is dominated by herding (or ``fashion'') effects. In this case, the individual Slutsky matrix is no longer symmetric, even for fully rational agents. The vicinity of the transition features a peak in asymmetry. 
\end{abstract}

\maketitle


\section*{\label{sec:intro}Introduction}
Economic theory still appears to be wedded to the strict \textit{Homo economicus} paradigm, despite numerous well established facts that contradict its very essence. In most models, economic agents are assumed to be fully rational, isolated, and homogeneous optimizers, capable of maximizing utility efficiently and instantaneously, at odds with many behavioral experiments and common sense. Even though boundedly rational \cite{simon1972theories,arthur1994inductive,gigerenzer2002bounded} or event irrational agents could behave collectively as a unique representative rational agent, this is very much a leap of faith, specially in the presence of strong interactions that can generate trends, fads and fashions, herding or mass panics, to name a few -- see e.g.~\cite{brockdurlauf2001,Kirman2010,Bouchaud2013,morelli2020confidence}. Empirical data also reveal very strong, Pareto heterogeneities in wealth, firm sizes, market capitalisation, etc.~\cite{bouchaud2001power,axtell2001zipf,farmer2008power,gabaix2009power} that may considerably affect aggregate quantities: this is Gabaix's ``granularity hypothesis''~\cite{gabaix2011granularity}.\footnote{Heterogenous agents are progressively (but only partly) accounted for in recent macroeconomic models, see e.g. \cite{kaplan2018monetary}.}  Other effects, such as habit formation and hysteresis can also play a major role in the behaviour of consumers, firms and investors alike~\cite{abel1990asset,campbell1999habit,carroll2000saving, moran2020by}.

 


The Slutsky matrix~\cite{slutsky1915}, central in consumer choice theory, stands as a textbook example of a mathematically sound economic construction that most likely lacks realism. It describes the response of consumption of good $i$ to a change of price of good $j$. From a theoretical standpoint, the Slutsky equation has been extensively investigated, see e.g.~\cite{Brown1972,Samuelson1974,Barten1982}, focusing notably on its geometrical interpretation~\cite{Grandville1989}. Under the classical assumptions mentioned above, very clear conclusions can be drawn: the Slutsky matrix is \textit{symmetric}, \textit{negative semi-definite}, and satisfies \textit{``homogeneity''} -- meaning the price vector of the products considered is an eigenvector of the matrix associated to a zero eigenvalue. 

Comparatively very little has been done from the empirical perspective. Looking at private consumption in the pre- and post-WWII periods in the US \cite{fuleihan1968empirical} and the Netherlands \cite{Barten1967}, statistical tests have ruled in favor of negative semi-definiteness but remained rather inconclusive regarding symmetry and homogeneity. Other studies have supported the symmetry condition \cite{Berndt1977,blundell1993we}, or found that it is satisfied for single households and violated for couples, but in a way that can be accounted for \cite{Browning1998}. Despite these conflicting conclusions, the Slutsky matrix and its associated conditions still stand at the core of consumer choice theory, or as nicely put by Barten~\cite{Barten1967}: \emph{``Such a situation is not uncommon in economics [...]. Many statements are generally accepted as being a true description of behaviour without a thorough questioning of their empirical validity.''} For further criticism of Slutsky theory and a particular focus on how income distribution changes are not taken into account, see~\cite{Mirowski1998}. In response to the inconclusiveness of empirical tests, some authors have shown that there is in fact no reason for symmetry to hold when taking into account the fact that consumers are many \cite{Diewert1977}, non-cooperative households \cite{lechene2011noncooperative}, or bounded rationality in the form of limited attention \cite{gabaix2014sparsity}. More recently, there have also been some attempts at taking into account the somewhat unrealistic nature of the assumptions behind the classical Slutsky equation, with the idea of providing a measure of some loosely defined rationality based on the nearness between the theoretical prediction and estimates constructed from observed demand functions \cite{aguiar2017slutsky,aguiar2018classifying}.

In this paper, we introduce a rigorous mathematical framework inspired by statistical physics that is able to accommodate bounded rationality, population heterogeneity, and interactions between agents, with in particular herding tendencies. We investigate how the inclusion of such behavioral features impacts the classical properties of the Slutsky matrix. One of our central results is a novel representation of the Slutsky matrix in terms of the consumption \textit{fluctuations} for fixed prices and budgets, instead of the traditional \textit{response} to a change of prices and budgets. Using this language, the symmetry property of the Slutsky matrix for a unique, boundedly rational agent becomes quite transparent. However, we show that including interactions between agents can and does affect this property -- although all eigenvalues still retain a non-positive real part.      

The model we propose to account for interactions between consumer choices turns out to be very interesting in its own right. We show that beyond a certain interaction strength, the choice of consumers \textit{condenses} onto a limited set of products, that become desirable just because others find it desirable, in a self-fulfilling fashion. We determine the boundaries of this condensed regime in parameter space. It turns out that the Slutsky matrix is maximally non-symmetric close to such boundaries.   

The paper is organized as follows. In Section~\ref{sec:slutsky}, we briefly introduce classical consumer choice theory and the Slutsky equation. Its generalization to bounded rationality and interacting agents is given in Section \ref{sec:Bounded_rationality}, where the fluctuation-response formulation of the problem is also derived. Two possible definitions of the aggregate Slutsky matrix are also discussed. In Section~\ref{sec:Animal_spirits}, we propose an interacting model and study the condensation of consumer choices. The consequences of this collective effect on the Slutsky matrix and its properties is then presented in Section~\ref{sec:Interacting_Slutsky}. The equivalence of the thermodynamic ensembles used and possible out-of-equilibrium effects are discussed in Section~\ref{sec:Discussion}. We finally summarize our findings and discuss possible future work in Section~\ref{sec:Conclusion}.

\section{\label{sec:slutsky} Consumer choice theory}

Consumer choice theory is based on the idea that, for a given bundle of $M$ goods with prices $\mathbf{p} \in \mathbb{R}_+^M$, agents choose a basket $\mathbf{x}$ to maximize their utility function $u(\mathbf{x})$, while subject to the constraint $\mathbf{p} \cdot \mathbf{x} = w$, where $w$ is the consumption budget. 

The utility function must then satisfy some elementary conditions. In particular, it is taken to be an increasing function of the quantity of goods, i.e.~more is always better and there is no satiation. While the utility must therefore have a positive first derivative, its rate of increase should be a decreasing function of the absolute quantity of goods. In other words, utility is postulated to be concave, i.e.~the marginal utility gain diminishes as goods are accumulated.

Assuming that the agent is fully rational, the optimal basket that maximizes the agent's satisfaction $\mathbf{x}^*$ is then simply given by
\begin{equation}
\label{argmax}
    \mathbf{x}^* = \underset {\mathbf{x} \geq 0}{\operatorname {argmax} } [u(\mathbf{x})\, | \, \mathbf{p} \cdot \mathbf{x} = w]\, .
\end{equation}
Conventionally one defines the \textit{Marshallian} demand $\mathbf{x}(\mathbf{p},w)$ which corresponds to that obtained from solving problem \eqref{argmax}, and the \textit{Hicksian} demand $\mathbf{h}(\mathbf{p},u)$ defined as the demand that minimizes the expenditure $e(\mathbf{p},u)$ for a fixed utility level $u$~\cite{Mas1995}. 
Setting $\mathbf{h}( \mathbf{p}, u) = \mathbf{x}^*(\mathbf{p}, e( \mathbf{p}, u))$ and differentiating, one obtains the Slutsky equation describing the change in consumption of good $i$ following a change in the price of good $j$
\begin{equation}
    \frac{\partial x_i}{\partial p_j}=\frac{\partial h_i}{\partial p_j}-x_j\frac{\partial x_i}{\partial w}.
    \label{equ:Slutsky_equ}
\end{equation}
Changes $\delta \mathbf{x}$ in the optimal basket's composition in response to a price change $\delta \mathbf{p}$ can thus be separated into two contributions: the \textit{substitution effect} (first term in the RHS of Eq.~\eqref{equ:Slutsky_equ}), describing how consumption is impacted by changes in relative prices of goods, and the \textit{income effect} (second term in the RHS of Eq.~\eqref{equ:Slutsky_equ}), expressing the impact of changes in purchasing power. The substitution effect is often described in terms of the Slutsky matrix $\mathbf{S}$, with entries therefore defined as
\begin{equation}
    S_{ij} := \frac{\partial h_i}{\partial p_j} = \frac{\partial x_i}{\partial p_j} + x_j \frac{\partial x_i}{\partial w}.
    \label{equ:Slutsky_mat}
\end{equation}
Provided that the utility function is sufficiently regular, $\mathbf{S}$ can then be shown to be symmetric, negative semi-definite, and equal to the Hessian of the expenditure function. In practice, the Slutsky matrix cannot be observed directly, but can be estimated as the other two terms in Eq.~\eqref{equ:Slutsky_mat} should be accessible empirically. In the following, we will give an alternative theoretical expression for the Slutsky matrix in terms of consumption fluctuations.

\section{Bounded Rationality}
\label{sec:Bounded_rationality}

As described in the introduction, taking agents to be perfect optimizers seems to be an unrealistic assumption in most contexts. 
There are several ways to relax such an assumption. One is that agents have a limited attention and cannot process all the information accessible to them, see e.g. \cite{gabaix2014sparsity} and refs. therein. As argued by Gabaix, this may effectively lead to \textit{perceived} prices that differ from real prices. This in turn affects the symmetry of the Slutsky matrix. 

Another traditional line of thought in the literature on choice theory is to replace the deterministic utility optimization prescription by a stochastic choice rule, such that the stationary choice distribution is given by a Boltzmann-Gibbs measure (see below) with an \textit{intensity of choice} parameter that allows one to interpolate between full rationality and completely random choices -- for a classical review, see \cite{anderson1992discrete}; see also \cite{Bouchaud2013} for more recent developments. This is the prescription we will adopt in the rest of the paper; as we shall see this allows us to describe a rather wide range of phenomena while ensuring mathematical tractability of the model.

\subsection{A Single Agent}
Formally, considering first a single agent, we postulate that the probability density for selecting the basket of goods $\mathbf{x}$ is given by
\begin{equation}
    P(\mathbf{x}) = 
    \begin{cases}
    \frac{1}{Z} \e^{\beta u(\mathbf{x})} \quad &\text{if } \mathbf{p}\cdot \mathbf{x} = w,\\
    0 \quad &\text{otherwise},
    \end{cases}
    \label{equ:canonical_distrib}
\end{equation}
where $u(\mathbf{x})$ is the utility function and the parameter $\beta$ is known as the intensity of choice (or the \textit{inverse temperature} in physics parlance). Taking $\beta \to \infty$, one recovers the fully rational case previously introduced and equation~\eqref{argmax}, while $\beta = 0$ means the agent picks any basket satisfying the constraint with equal probability. The normalization factor $Z$ is known in statistical physics as the partition function, which, given the hard budget constraint enforced here, writes
\begin{equation}
    Z = \int_+ \dd\mathbf{x} \, \e^{\beta u(\mathbf{x}) } \delta(\mathbf{p}\cdot\mathbf{x} - w),
\end{equation}
where $\int_+$ means that we integrate over non-negative baskets $\mathbf{x} \geq 0$. For finite values of $\beta$, the basket $\mathbf{x}$ must now be statistically described, since different realizations of the system will lead to different outcomes. A suited definition of the Slutsky matrix must therefore be considered. Here, we propose to replace all $x_i$ by their averages $\langle x_i \rangle$,
\begin{equation}
    S_{ij} := \frac{\partial}{\partial p_j} \langle x_i \rangle + \langle x_j \rangle \frac{\partial}{\partial w} \langle x_i \rangle,
    \label{equ:Slutsky_canonical}
\end{equation}
with angular brackets referring to an average over the distribution given in Eq.~\eqref{equ:canonical_distrib}, to wit
\[ 
\langle O(\mathbf{x}) \rangle := \frac1Z 
\int_+ \dd\mathbf{x} \, O(\mathbf{x}) \, \e^{\beta u(\mathbf{x})} \delta(\mathbf{p}\cdot\mathbf{x} - w).
\]
Note that in the limit $\beta \to \infty$, one recovers the standard Slutsky matrix since $\langle x_i \rangle \to x_i^*$.

Our set up of the problem allows us to draw several analogies with the statistical mechanics foundation of thermodynamics. For example, the strict application of the budget constraint  $\delta(\mathbf{p}\cdot\mathbf{x} - w)$ is reminiscent of the so-called  \textit{canonical} ensemble, where the conservation of the number of particles in the system is strictly enforced. We will see later that one can also work in the analogue of the \textit{grand-canonical} ensemble where the budget constraint is only enforced on average. This eases some analytical calculations while being equivalent to the canonical ensemble in some limits (for example when the number of goods is large). One can argue that in some cases, allowing the budget to fluctuate (due to loans for example) can be realistic as well.

\subsection{A Fluctuation-Response Relation}

More interestingly, statistical mechanics also provides relations between the response of certain quantities to external perturbations to spontaneous fluctuations of these quantities in the absence of perturbations. These relations can be established using the derivatives of the partition function, assuming an equilibrium state has indeed been reached. In particular, the Slutsky matrix can indeed be expressed in terms of other correlations, as was first mentioned in one of the present authors (JPB) contribution to \cite{ekeland2011}.  In Appendix~\ref{appendix:SingleAgentSlutsky}, we derive the following ``fluctuation-response'' formula in the single agent case:
\begin{equation}
    S_{ij} = -\Gamma \langle x_i x_j \rangle_c - \partial_w \langle x_i x_j \rangle_c,\label{eq:symslutskyc}
\end{equation}
with $\Gamma = \partial_w \log Z$ and $\langle x_i x_j \rangle_c := \langle x_i x_j \rangle - \langle x_i \rangle \langle x_j \rangle$. Equation~\eqref{eq:symslutskyc} is manifestly symmetric in $i$,$j$. It shows that even with  bounded rationality, the Slutsky matrix is still symmetric, for any value of $\beta$, not only in the rational limit $\beta \to \infty$. Hence, \textit{symmetry of the Slutsky matrix may not be used as a proof of the rationality of economic agents}, as was argued in \cite{ekeland2011} for example. 

Our fluctuation formula Eq. \eqref{eq:symslutskyc} is also interesting from an econometric standpoint, as it provides a way to measure the Slutsky matrix without varying prices. Measuring response quantities from equilibrium correlations is in fact commonly used in statistical mechanics, through what is referred to as \textit{fluctuation-dissipation} relations \cite{marconi2008fluctuation}, or, in a restricted context, to Einstein's relation relating mobility to diffusion for Brownian particles.

One can go one step further and eliminate all derivatives from Eq. \eqref{eq:symslutskyc}, provided the utility function is known. One finds (see Appendix~\ref{appendix:FDR})
\begin{equation}
\begin{aligned}
    &\frac{\partial}{\partial w} \langle x_i x_j \rangle_c = \frac{\beta}{p_k} \Big[ \Big\langle x_i x_j \frac{\partial u}{\partial x_k} \Big\rangle + 2\langle x_i \rangle \langle x_j \rangle \Big\langle \frac{\partial u}{\partial x_k} \Big\rangle \\
    &-  \langle x_i x_j \rangle \Big\langle \frac{\partial u}{\partial x_k} \Big\rangle- \langle x_i \rangle \Big\langle x_j \frac{\partial u}{\partial x_k} \Big\rangle - \langle x_j \rangle \Big\langle x_i \frac{\partial u}{\partial x_k} \Big\rangle\Big],
\end{aligned}
\end{equation}
and 
\begin{equation}
    \Gamma = \frac{\beta}{p_k} \Big\langle \frac{\partial u}{\partial x_k} \Big \rangle,
    \label{equ:Gamma_FDT_1}
\end{equation}
both equations being valid for an arbitrary choice of $k$.

\subsection{Many Agents}
\label{sec:many_agents}
In practice, agents make correlated choices and we must adapt our formalism to treat interactions. We thus consider $N$ agents, indexed by $\alpha = 1,\dots,N$, and denote $U(\{\mathbf{x}^\alpha\})$ the aggregate utility of all the agents. Interactions mean that $U$ cannot in general be written as a sum of individual utility functions $u^\alpha$. The probability that the $N$ agents choose a bundle of goods $\{\mathbf{x}^\alpha\}$ is proportional to $\exp(\beta U)/Z_N$, where the aggregate partition function $Z_N$  writes
\begin{equation}
    Z_N = \int_+ \bigg( \prod_{\alpha=1}^N \prod_{i=1}^M \dd x_i^\alpha \bigg) \, \e^{\beta U(\{ \mathbf{x}^\alpha \})} \prod_{\alpha = 1}^N \delta(\mathbf{p} \cdot \mathbf{x}^\alpha - w^\alpha).
    \label{equ:Z_general}
\end{equation}
Here, we integrate over the $M \times N$ degrees of freedom, i.e. the quantities $x_i^\alpha$ of good $i$ consumed by agent $\alpha$, while enforcing that all agents respect their own specific budget $w^\alpha$.\footnote{It should be noted that with this partition function we implicitly assume that all agents work at improving the aggregate utility of the system, which may not be the case in reality. For a complete discussion addressing the possible differences when considering purely individualistic agents, see Section~\ref{sec:global_vs_individual}.} Analytically computing this partition function is usually very difficult due to the product of Dirac~$\delta$ distributions in the integrand. Depending on the form of the utility, one might need to slightly relax the budget constraints. As mentioned above, a way to do this is to move to the so-called \textit{grand-canonical} partition function $\mathcal{Z}_N$, defined as
\begin{equation}
    \mathcal{Z}_N = \int_+ \bigg( \prod_{\alpha=1}^N \prod_{i=1}^M \dd x_i^\alpha \bigg) \, \e^{\beta [U(\{ \mathbf{x}^\alpha \}) - \sum_\alpha \mu^\alpha \mathbf{p} \cdot \mathbf{x}^\alpha]},
\end{equation}
where the $\mu^\alpha$, known in physics as ``chemical potentials'', are fixed by enforcing that budgets are satisfied on average, i.e. $\mathbf{p} \cdot \langle \mathbf{x}^\alpha \rangle  = w^\alpha$, $\forall \alpha$. In general, the two partition functions are not equivalent; however, many quantities calculated from the two ensembles become identical in the large $M$ limit and/or in the rational limit $\beta \to \infty$, see Section \ref{sec:equivalence} for a detailed discussion.

Consistent with the single agent definition, we take the individual Slutsky matrix of agent $\alpha$ to be given by
\begin{equation}
    S_{ij}^\alpha := \frac{\partial}{\partial p_j} \langle x_i^\alpha \rangle + \langle x_j^\alpha \rangle \frac{\partial}{\partial w^\alpha} \langle x_i^\alpha \rangle,
    \label{equ:Slutsky_mat_gen}
\end{equation}
where we have assumed that prices are the same for all agents. Note that we have also taken a uniform system-wide rationality parameter $\beta$, although a generalization to different $\beta^\alpha$'s is possible and would be an interesting extension of our work.

As in the single agent case, the partition function allows one to derive a fluctuation-response expression for the Slutsky matrix in terms of correlations. Equation~\eqref{equ:Slutsky_mat_gen} may be rewritten (see Appendix~\ref{appendix:MultiAgentSlutsky}) in what will be referred to as its ``thermodynamic'' form
\begin{equation}
\begin{aligned}
   S_{ij}^\alpha = - \sum_\gamma \Big[ &\Gamma_\gamma \langle x_i^\alpha x_j^\gamma \rangle_c + \frac{\partial}{\partial w^\gamma} \langle x_i^\alpha x_j^\gamma \rangle_c \\
   & + (1-\delta_{\alpha \gamma}) \langle x_j^\gamma \rangle \frac{\partial}{\partial w^\gamma} \langle  x_i^\alpha \rangle \Big],
\end{aligned}
\label{equ:Slutsky_noderivatives}
\end{equation}
with $\Gamma_\gamma = \partial_{w^\gamma} \log Z_N$. This is the central theoretical result of this paper, which, to the best of our knowledge, is new. Although its derivation makes explicit use of the Boltzmann-Gibbs distribution, such an expression holds more generally in the near-rational limit $\beta \to \infty$ where small deviations around the rational solution are always Gaussian, see Section \ref{sec:near_rational} below.

Importantly, unlike in the single-agent case, Eq. \eqref{equ:Slutsky_noderivatives} does not allow one to infer anything general about the symmetry of the Slutsky matrices. In the case where interactions between agents are negligible, i.e. when $U$ is the sum of individual utility functions, correlations between agents are zero whenever $\gamma \neq \alpha$ and we recover the single agent expression, as expected.

Assuming the utility function is known, one can again go further and express derivatives with respect to budgets $w^\gamma$ as a function of some correlations. For our problem we find a generalization of the formula obtained for a single agent in the canonical ensemble (see Appendix~\ref{appendix:FDR})
\begin{equation}
\begin{aligned}
    &\frac{\partial}{\partial w^\gamma} \langle x_i^\alpha x_j^\gamma \rangle_c = \frac{\beta}{p_k} \Big[ \Big\langle x_i^\alpha x_j^\gamma \frac{\partial U}{\partial x_k^\gamma} \Big\rangle 
    + 2\langle x_i^\alpha \rangle \langle x_j^\gamma \rangle \Big\langle \frac{\partial U}{\partial x_k^\gamma} \Big\rangle
    \\ &- \langle x_i^\alpha x_j^\gamma \rangle \Big\langle \frac{\partial U}{\partial x_k^\gamma} \Big\rangle -
    \langle x_i^\alpha  \rangle \Big\langle x_j^\gamma \frac{\partial U}{\partial x_k^\gamma} \Big\rangle  -
    \langle x_j^\gamma \rangle \Big\langle x_i^\alpha \frac{\partial U}{\partial x_k^\gamma} \Big\rangle \Big]
    ,
\end{aligned}
\end{equation}
as well as
\begin{equation}
    \Gamma_\gamma = \frac{\beta}{p_k} \Big\langle \frac{\partial U}{\partial x_k^\gamma} \Big \rangle,
    \label{equ:Gamma_FDT}
\end{equation}
both again valid for any  good $k$. From these expressions and given a utility function $U$, all terms in Eq.~\eqref{equ:Slutsky_noderivatives} can be computed in principle, at least numerically. Bear in mind, however, that these relations are only valid if the system has reached equilibrium, which might take a very long time, for example near a phase transition point.

\subsection{Aggregate Slutsky Matrices}

There are {\it a priori} two possible definitions of the Slutsky matrix at the aggregate level. One is simply to take the average over all agents of the individual Slutsky matrices, to wit
\begin{equation}
    \overline{S}_{ij} := \frac1N \sum_\alpha S_{ij}^\alpha.
\end{equation} 
However, empirical measurements often rely on estimates at the aggregate level. In that case, a better suited definition uses aggregate consumption:
\begin{equation}
    \mathcal{S}_{ij} := \frac{\partial}{\partial p_j} \langle \overline{x}_i \rangle + \langle \overline{x}_j \rangle \frac{\partial}{\partial \overline{w}} \langle \overline{x}_i \rangle,
\end{equation}
with overlines indicating averaging over agents e.g.
\[
\overline{x}_i := \frac{1}{N} \sum_{\alpha} x_i^\alpha,
\]
and similarly for the average budget $\overline{w}$. In that case, as shown in Appendix \ref{appendix:Slutsky_aggregate}, the thermodynamic expression becomes
\begin{equation}
\begin{aligned}
        \mathcal{S}_{ij} =  - \frac1N \sum_{\alpha, \gamma} \bigg[ &\Gamma_\gamma \langle x_i^\alpha x_j^\gamma\rangle_c + \frac{\partial}{\partial w^\gamma} \langle x_i^\alpha x_j^\gamma \rangle_c \\
        & +  \left(\langle x_j^\gamma \rangle  - \kappa^\gamma {\overline{x}_j} \right) \frac{\partial}{\partial w^\gamma} \langle x_i^\alpha \rangle   \bigg],
\end{aligned}
\end{equation}
with the possibly heterogeneous factor
\[ 
\kappa^\gamma := \frac{\partial {w}^\gamma}{\partial \overline{w}}.
\]
Clearly, if consumption is proportional to wages and if all wages scale with the average wage, i.e. $w^\gamma = \kappa^\gamma \overline{w}$, then $\langle x_j^\gamma \rangle = \kappa^\gamma \overline{x}_j$, and there is no contribution from the last term. In this case, we find an expression very close to the single-agent case. More generally, $\mathcal{S}_{ij}$ has no reason to be symmetric, except when all agents are identical. In such a case, even in the presence of interactions and for bounded rationality ($\beta < +\infty$), $\mathcal{S}_{ij}$ is  always symmetric, whereas $\overline{S}_{ij}$ is not, as we in the next sections. 

\subsection{Near-Rational Limit $\beta \to \infty$}
\label{sec:near_rational}

In order to simplify the problem and get some intuition, we place ourselves in the near-rational case where $\beta \to \infty$. This corresponds to the low temperature case in a physical system, where one can expect that all relevant configurations $\{ \mathbf{x}^\alpha \}$ are close to the optimal configuration $\{\mathbf{x}^{\alpha *} \}$ that maximizes the global utility function $U$ subject to budget constraints. We thus write $\delta x_i^\alpha := x_i^\alpha -x_i^{\alpha *}$ and Taylor-expand the utility function to second order, resulting in
\begin{equation}
    U(\{\mathbf{x}^\alpha\}) \approx U^* + \frac{1}{2} \{\delta \mathbf{x}^\alpha \}^\top \mathbf{H} \{\delta \mathbf{x}^\alpha \} + O(\delta x^3),
\end{equation}
with $\mathbf{H}$ the $(M\times N) \times (M\times N)$ Hessian of the system evaluated at the maximum of the utility function, for a given set of budget constraints $w^\alpha$. Here we only consider deviations $\{\delta \mathbf{x}^\alpha \}$ that all satisfy the budget constraints, so that the first derivative terms $U'$ are zero as we expand around a maximum along all directions but one. Calculating the partition function and correlations then simply amounts to computing Gaussian integrals, although the budget constraints must still be handled with care.\footnote{As noted above, for large $\beta$ all fluctuations are small and Gaussian, so in that limit our results are actually more general than the Boltzmann-Gibbs model on which they are based.} By taking the Fourier representation of the Dirac $\delta$ (see Appendix \ref{appendix:Gaussian_correls}), we finally find, to leading order in $\beta^{-1}$:
\begin{equation}
\begin{aligned}
    {C}_{ij}^{\alpha \gamma} :=& \langle x_i^\alpha  x_j^\gamma \rangle_c =  -\frac{1}{\beta} \bigg[ (\mathbf{H}^{-1})_{ij}^{\alpha \gamma} \\ 
    & - \sum_{\eta,\nu,k,\ell} (\mathbf{H}^{-1})_{ik}^{\alpha \eta} p_k (\mathbf{G}^{-1})^{\eta \nu} p_\ell (\mathbf{H}^{-1})_{\ell j}^{\nu \gamma}  \bigg],
\end{aligned}
\label{equ:correls}
\end{equation}
with the $N \times N$ matrix $\mathbf{G}$ defined as
\begin{equation}
    G^{\eta \nu} = \sum_{i,j} p_i (\mathbf{H}^{-1})_{ij}^{\eta \nu} p_j,
\end{equation}
and where the second term in the RHS of Eq.~\eqref{equ:correls} is the result of the constraint being applied. 

Let us first illustrate this formula in the $N=1$ case. In the canonical ensemble and large $\beta$ regime, one has
\[
\Big\langle \frac{\partial u}{\partial x_k} \Big \rangle = \lambda p_k + O(\beta^{-2}),
\] 
where $\lambda$ is the Lagrange parameter enforcing that the budget constraint is satisfied. Hence the first term Eq.~\eqref{eq:symslutskyc} remains finite since $\Gamma$ diverges as $\beta$ when $\langle x_i x_j \rangle_c$ tends to zero as $\beta^{-1}$. The second term, on the other hand, tends to zero at least as $\beta^{-1}$. It then follows that for a single and near-rational agent,
\begin{equation}
    \mathbf{S} =  \lambda \left[\mathbf{H}^{-1} - \mathbf{u} \mathbf{u}^\top\right] + O(\beta^{-1}), \quad \mathbf{u} = \frac{\mathbf{H}^{-1} \mathbf{p}}{\sqrt{\mathbf{p}^\top \mathbf{H}^{-1} \mathbf{p}}}.
\end{equation}
This is the classic expression for the Slutsky matrix, which our framework therefore allows  to recover in the corresponding limit, with corrections in $\beta^{-1}$ that can be computed. Since the Hessian is both symmetric and negative semi-definite at a utility maximum, we recover the classic properties of the Slutsky matrix. ``Homogeneity'' is also easily recovered by checking that multiplying the matrix by $\mathbf{p}$ indeed gives a zero eigenvalue.

When $N > 1$, we will need to specify the utility function $U$ to make the final result more explicit. Keeping $U$ fully general and taking the limit $\beta \to \infty$ only allows one to simplify the general expression Eq. \eqref{equ:Slutsky_noderivatives} to
\begin{equation}
\begin{aligned}
    S_{ij}^\alpha = - \beta \sum_\gamma \lambda_\gamma C_{ij}^{\alpha\gamma} - \sum_{\gamma \neq \alpha}  \frac{\langle x_j^\gamma \rangle}{p_j} \sum_{k,\nu} H_{ik}^{\alpha \nu} C_{ik}^{\alpha \nu} + O(\beta^{-1})
    \label{equ:Slutsky_T=0},
\end{aligned}
\end{equation}
where we have used Eq.~\eqref{eq:partialwx} and expanded $\partial_x U$ to first order in $\delta x$. A similar expression can be derived for the aggregate Slutsky matrix $\mathcal{S}_{ij}$ as well. Remember that $\mathbf{C}$ is of order $\beta^{-1}$ (see Eq.~\eqref{equ:correls}) so $S_{ij}^\alpha$ is well behaved in the large $\beta$ limit. Although the final expression is not transparent, it is clear that $S_{ij}^\alpha$ has no reason to be always symmetric, except when agents' choices are uncorrelated, in which case $C_{ij}^{\alpha\gamma}=0$ whenever $\alpha \neq \gamma$ and $C_{ij}^{\alpha\alpha}$ is symmetric by construction. We will now turn to an explicit model with herding, where the asymmetry of Slutsky matrix can be made apparent. 

\section{Animal spirits}
\label{sec:Animal_spirits}

As argued in the introduction, both anecdotal evidence about fads and fashion and more serious scientific studies point to the fact that agents' choices can be strongly influenced by the choice of others (see e.g. \cite{veblen,bass1969new,brockdurlauf2001,salganik2006experimental,gordon2009discrete, Bouchaud2013}), an effect also known as ``keeping up with the Joneses'' \cite{campbell1999habit, stracca2005keeping}, see also \cite{morelli2020confidence}. We now propose a family of models which account for interactions between boundedly rational agents. Interestingly, these models lead, in some regions of parameters, to ``concentration'' (or ``condensation'') of choices, much in the spirit of the model proposed by Borghesi and Bouchaud \cite{Borghesi2007} (see also \cite{lucas2022non} for a recent extension). As we shall see, close to the concentration transition, the non-symmetric contribution to the Slutsky matrix reaches a maximum. 

\begin{figure}
    \centering
    \includegraphics[width=\linewidth]{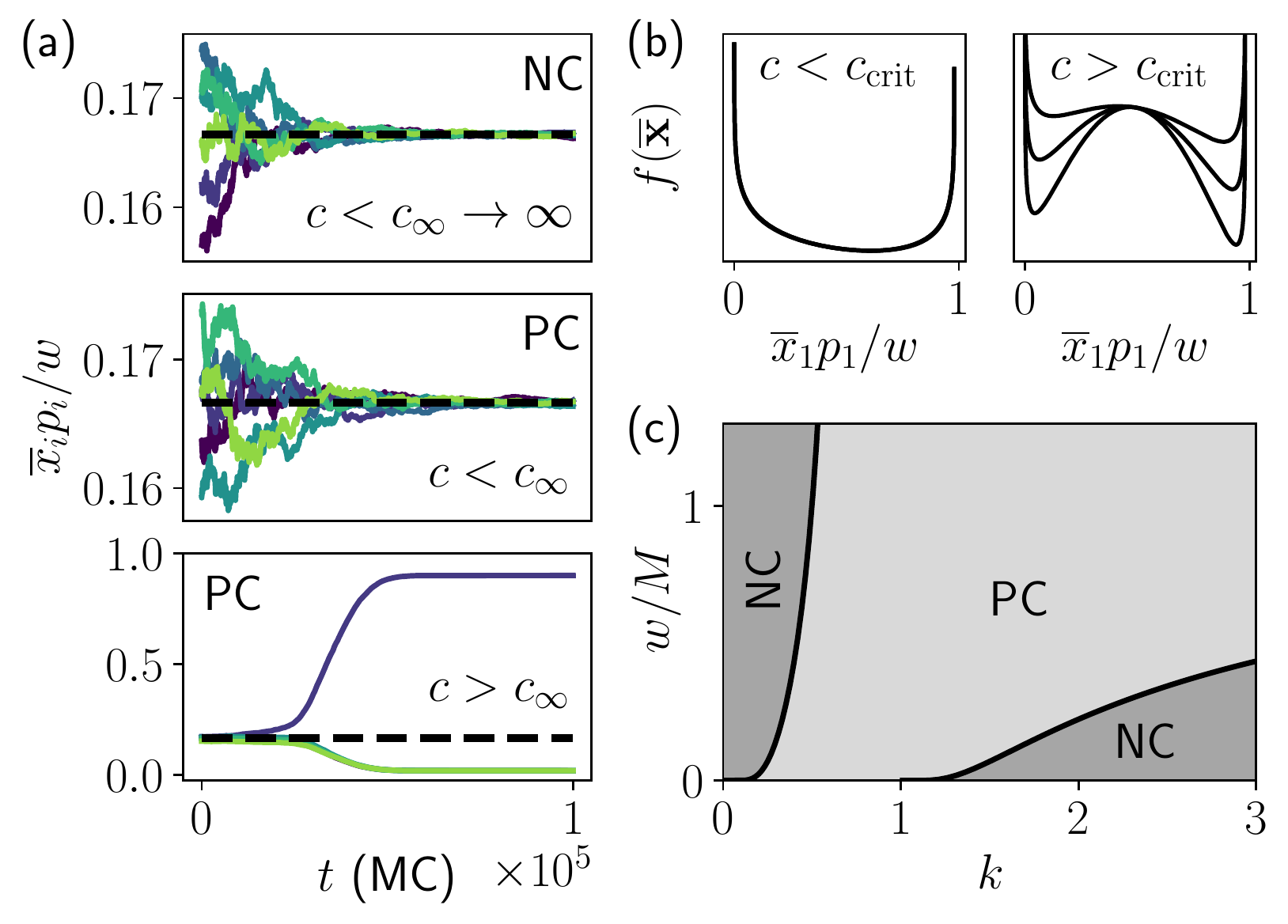}
    \caption{(a) Monte Carlo simulations at $\beta \to \infty$ for $M = 6$ products, $N=256$ identical agents, $k=2$ and $p_i = a_i = 1$, compared to the non-condensed solution (dashed line). Top: Non-Condensed (NC) region where the value of $c_\infty$ diverges [$w = 0.5$, $c=1$]. Middle: Possible Concentration (PC) region for $c < c_\infty$ [$w = 10$, $c = 0.01$]. Bottom: PC region for $c > c_\infty$ [$w = 10$, $c = 0.1$], the system concentrates on one of the products and departs from the non-condensed solution. (b) Illustration of the free energy $f$ for some temperature $\beta$ in the PC phase for $M = 2$ products, $p_2 > p_1$ and $a_2=a_1$. Left: before the transition, $c < c_\mathrm{crit}$. Right: after the transition $c > c_\mathrm{crit}$, the deeper minima corresponding to increasing values of $c$. (c) Theoretical phase diagram of the NC and PC regions at $\beta \to \infty$, $a_i = p_i =1$.} 
    \label{fig:Transition_betainf}
\end{figure}

\subsection{Interactions and Herding}

In order to study imitation or fashion effects, we first consider agents with log-utilities and take the aggregate utility to simply be the sum of all agent-specific utilities,
\begin{equation}
    U(\{\mathbf{x}^\alpha\}) = \sum_{\alpha = 1}^N \sum_{i=1}^M a_i^\alpha \log x_i^\alpha.
    \label{equ:utility_global}
\end{equation}
where $a_i^\alpha$ describes the preference of agent $\alpha$ for good $i$. In the following, we assume that agents are homogeneous ($a_i^\alpha=a_i$, $\forall \alpha$) but interacting, by which we mean that the preference for good $i$ depends on how much good $i$ is consumed by other agents. Mathematically, we posit that the preference for goods increases with the $k$-th power of their average consumption:
\begin{equation}
    a_i \to a_i \left[1 + c (\overline{x}_i)^k \right],
\end{equation}
where we remind $\overline{x}_i$ is the average consumption of good $i$ (over all agents), and $c$ and $k$ are  non-negative parameters that describe the strength and nature of the interactions.\footnote{For an alternative specification, see Appendix \ref{appendix:Hamiltonian_U}.} The non-interacting case $c=0$ (or equivalently $k=0$, up to some rescaling of the $a_i)$ may be treated exactly in the canonical ensemble, i.e. strictly enforcing the budget constraint. As detailed in Appendix~\ref{appendix:heterogeneous}, the equilibrium configurations are given by
\begin{equation}
    \langle x_i^\alpha\rangle = \frac{w^\alpha}{p_i} \frac{1 + \beta a_i}{\sum_k(1+\beta a_k)},
\end{equation}
which will henceforth be referred to as the non-condensed or uniform solution. This solution  matches results from numerical experiments, as illustrated in the  two top panels of Figure~\ref{fig:Transition_betainf}(a) in the fully rational case, for example. For details about how numerical simulations have been performed, see Appendix \ref{appendix:mc}. The agent specific Slutsky matrices $S_{ij}^\alpha$ can then also be written explicitly from the original definition, verifying both symmetry and negative semi-definiteness for any $\beta$. In this non-interacting case, budget heterogeneities simply affect the magnitude of any given agent Slutsky matrix entries and are thus inconsequential for the properties of interest. 

Taking $k>0$ and increasing $c$, we expect the system to progressively depart from this solution and concentrate on some product(s) as $c\to \infty$, as observed in the numerical simulations shown in Fig.~\ref{fig:Transition_betainf}(a), bottom panel. To evaluate how this concentration occurs, we start by taking $w^\alpha = w, \; \forall \alpha$, that is a system of interacting but identical agents, both in wealth and preferences. As is often the case in statistical mechanics, the partition function cannot be computed exactly for interacting systems for general $N$, but can be more and more accurately approximated in the large $N$ limit. 

In our case, it is convenient to relax the budget constraint and to place ourselves in the grand-canonical ensemble where the budget constraint is only enforced on average. The procedure, detailed in Appendix \ref{appendix:interacting}, allows us to rewrite the grand-canonical partition function as an integral over the mean consumption vector $\overline{\mathbf{x}}$
\begin{equation}
    \mathcal{Z}_N = \int_{0}^\infty \dd \overline{\mathbf{x}} \, \e^{-N \beta f(\overline{\mathbf{x}})},
\end{equation}
where $f$ is usually called the ``free energy density'', here given by
\begin{widetext}
\begin{equation}
    \beta f(\overline{\mathbf{x}}) = \sum_i \big[ \beta \mu p_i \overline{x}_i - (1 + \beta a_i[1 + c(\overline{x}_i)^k])(1 + \log \overline{x}_i - \log(1+\beta a_i [1+c(\overline{x}_i)^k]) - \log \Gamma(1+\beta a_i [1+c(\overline{x}_i)^k]) \big] + o(1).
\end{equation}
\end{widetext}
with $\mu$ the ``chemical potential'' used to enforce the constraint on average, taken to be identical for all agents and $\Gamma(\cdot)$ is the Gamma function. Importantly, this grand-canonical description allows one to have a free energy density $f(\overline{\mathbf{x}})$ that is a sum over entirely decoupled goods. The only coupling is through the value of $\mu$. 

Given that $N \to \infty$, the shape of the free energy then completely determines the state of the system. Indeed, for large $N$ the partition function can be estimated using Laplace's method, such that the values $\overline{x}_i^* = \langle x_i \rangle$ that minimize the free energy are overwhelmingly more probable than any other values. Setting $c = 0$ and solving the set of equations $\frac{\partial f}{\partial \overline{x}_i} = 0$, one can for example check that the previously obtained solution is recovered.

As in statistical physics, we expect to identify the phase transition from the uniform to the \textit{condensed} (concentrated) phase where herding dominates \cite{Borghesi2007,Bouchaud2013}. This occurs when the single minimum in free energy associated to the non-condensed solution becomes a maximum while one or several new minima appear. Such a change of topology occurs for some value $c_\mathrm{crit}$, as illustrated in Fig.~\ref{fig:Transition_betainf}(b) in the case where there are only two products. For given values of $c$, the depth of the  minima in the concentrated phase will depend on the $a_i$ and $p_i$. The most favorable configuration (in this case having more of the least expensive of the two products, as here $a_i = 1, \, \forall i$) is associated to a lower free energy.

\subsection{Concentration for $\beta \to \infty$}

In order to find precisely when and how concentration occurs, we first study the  case of fully rational agents, $\beta \to \infty$. Carefully rescaling the free energy density, we find that an extremum is reached for the configurations $\overline{\mathbf{x}}^*$ satisfying
\begin{equation}
    a_i[1 + c(\overline{x}_i^*)^k (1+ k\log \overline{x}_i^*)] = \mu p_i \overline{x}_i^*,
    \label{equ:x_i_betainf}
\end{equation}
with the value of the chemical potential being such that
\begin{equation}
    \sum_i \overline{x}_i^* p_i = w.
\end{equation}
As previously described, the critical value $c_\mathrm{crit} = c_\infty$ where this transition occurs in the fully rational case can be found by looking at the Hessian of the free energy evaluated at the non-condensed solution. Doing so (see Appendix \ref{appendix:interacting}), and choosing for the sake of simplicity $a_i = p_i = 1$ for all agents, one finds
\begin{equation}
    \frac{1}{c_\infty} = \left( \frac{w}{M} \right)^k \left[ 2k - 1 + k(k-1) \log \left( \frac{w}{M} \right) \right].
\end{equation}
As long as the right hand side is strictly positive, there will therefore be a value of $c$ above which concentration occurs when $\beta \to \infty$, while if this is not the case concentration never occurs, regardless of the strength of interactions. The resulting theoretical phase diagram is shown in Fig.~\ref{fig:Transition_betainf}(c) for $k=2$, a value that we shall keep fixed henceforth. This phase diagram is in perfect agreement with our numerical simulations, as shown in Fig.~\ref{fig:Transition_betainf}(a). Note that this procedure can be repeated for different values of $a_i$ and $p_i$, leading qualitatively similar results. Since the Hessian of the free energy is diagonal in our case, the critical value $c_\infty$ would then correspond to the first change of sign of a diagonal element of the matrix.


\subsection{Finite $\beta$ Effects}

\begin{figure}
    \centering
    \includegraphics[width=\linewidth]{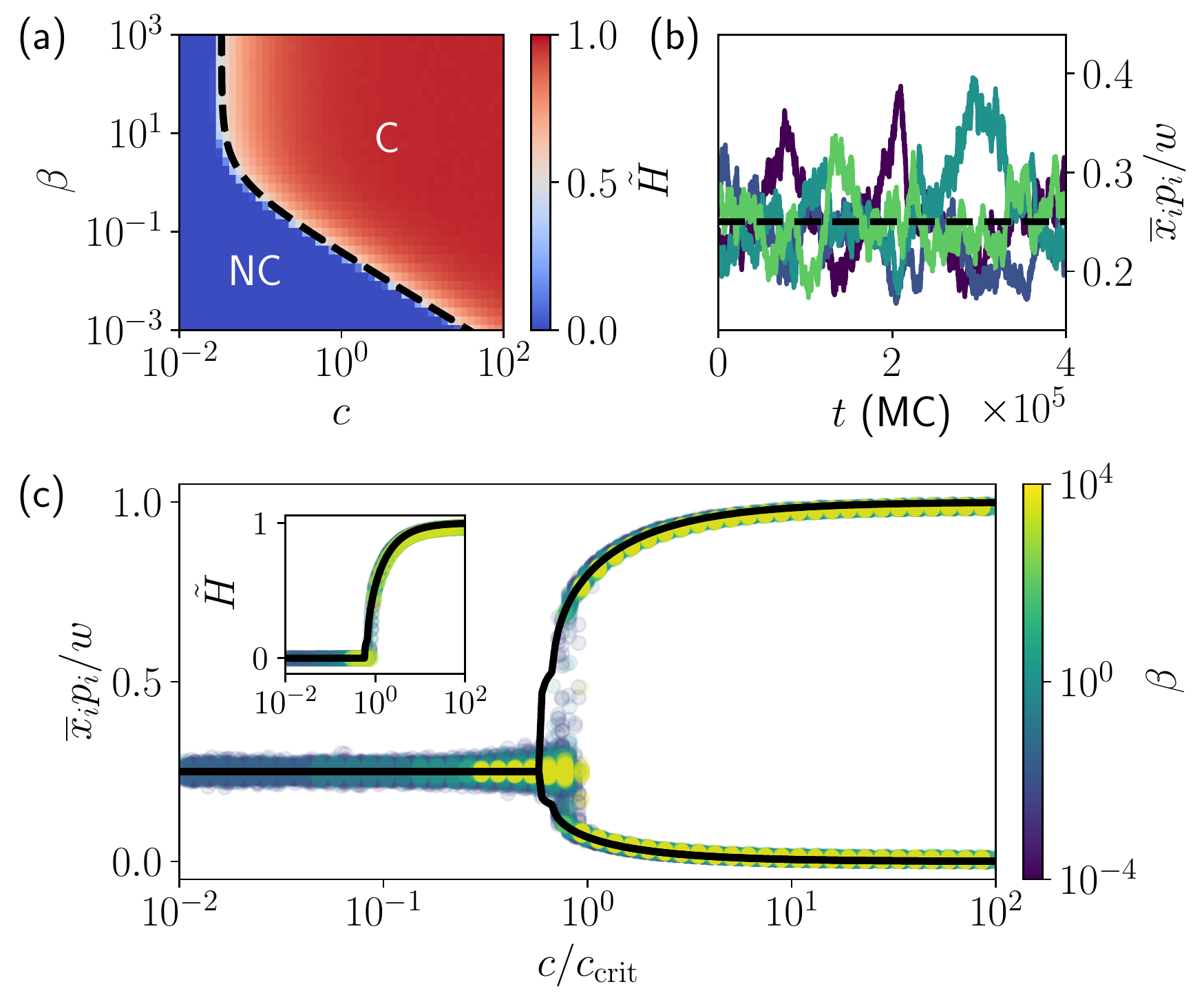}
    \caption{Effect of bounded rationality on concentration from numerical experiments for $M=4$ products, $N=64$ agents, $k=2$, $w=10$, $p_i=a_i=1$. (a) Herding phase diagram, with the normalized Herfindahl index $\tilde{H}$ calculated over the mean basket among agents, giving $\tilde{H} = 0$ in the Non-Condensed (NC) phase and quickly reaching $\tilde{H} = 1$ when condensation (C) occurs, indicating fashion dominated consumption. The critical line calculated analytically is shown by the dashed line, and perfectly matches numerical results. (b) Monte-Carlo dynamics of the mean basket for $\beta=1$, $c \approx c_\mathrm{crit}$, showing large fluctuations and switching behaviour. The non-condensed solution is shown by the dashed line. (c) Evolution of the rescaled average basket for different values of the rationality parameter $\beta$ as a function of $c$, collapsed using the analytical values of $c_\mathrm{crit}$ visible in the phase diagram. The analytical solution at $\beta \to \infty$ is shown as the continuous line. Inset: normalized Herfindahl, that clearly reveals the transition around $c_\mathrm{crit}$ for all values of $\beta$.}
    \label{fig:Transition_betafinite}
\end{figure}

Placing ourselves in the regions where condensation does occur in the fully rational limit, we now set out to understand how bounded rationality might alter the phase transition. In the general case, analytical expressions are difficult to obtain. However, numerically finding where the Hessian (which is still diagonal in $i,j$) loses stability for the non-condensed solution yields a semi-analytical critical line in $(c,\beta)$ space. It should be noted that this condition, explicitly given in Appendix \ref{appendix:interacting}, is  independent of the value of $\mu$, suggesting that the location of the transition is identical in the canonical and grand-canonical descriptions. 

Our analytical result can be compared to numerical simulations, for which the transition to fashion dominated consumption is identified by looking at the rescaled Herfindahl index
\begin{equation*}
    \tilde{H} = \frac{\sum_i (\overline{x}_i p_i/w)^2 - 1/M}{1-1/M}.
\end{equation*}
This index takes the values 0 and 1 in the uniform ($\overline{x}_i = w/(M p_i)$,  $\forall i$) and fully concentrated ($\overline{x}_i = w/p_i$ for one product and $=0$ for all others) cases respectively. As shown in Fig.~\ref{fig:Transition_betafinite}(a), the numerical phase diagram and the theoretical critical line match very well, despite the fact that the semi-analytical calculation is based on the grand-canonical ensemble, whereas numerical simulations strictly enforce the budget constraints for all agents. Furthermore, using the theoretical values for $c_\mathrm{crit}(\beta)$ to rescale the evolution of the average basket as a function of $c/c_\mathrm{crit}$ (as plotted in Fig.~\ref{fig:Transition_betafinite}(c)), we observe that the evolution of the mean basket and related quantities appears to be largely independent of $\beta$.

\section{Consequences on the Slutsky matrix}
\label{sec:Interacting_Slutsky}

\begin{figure}
    \centering
    \includegraphics[width=\linewidth]{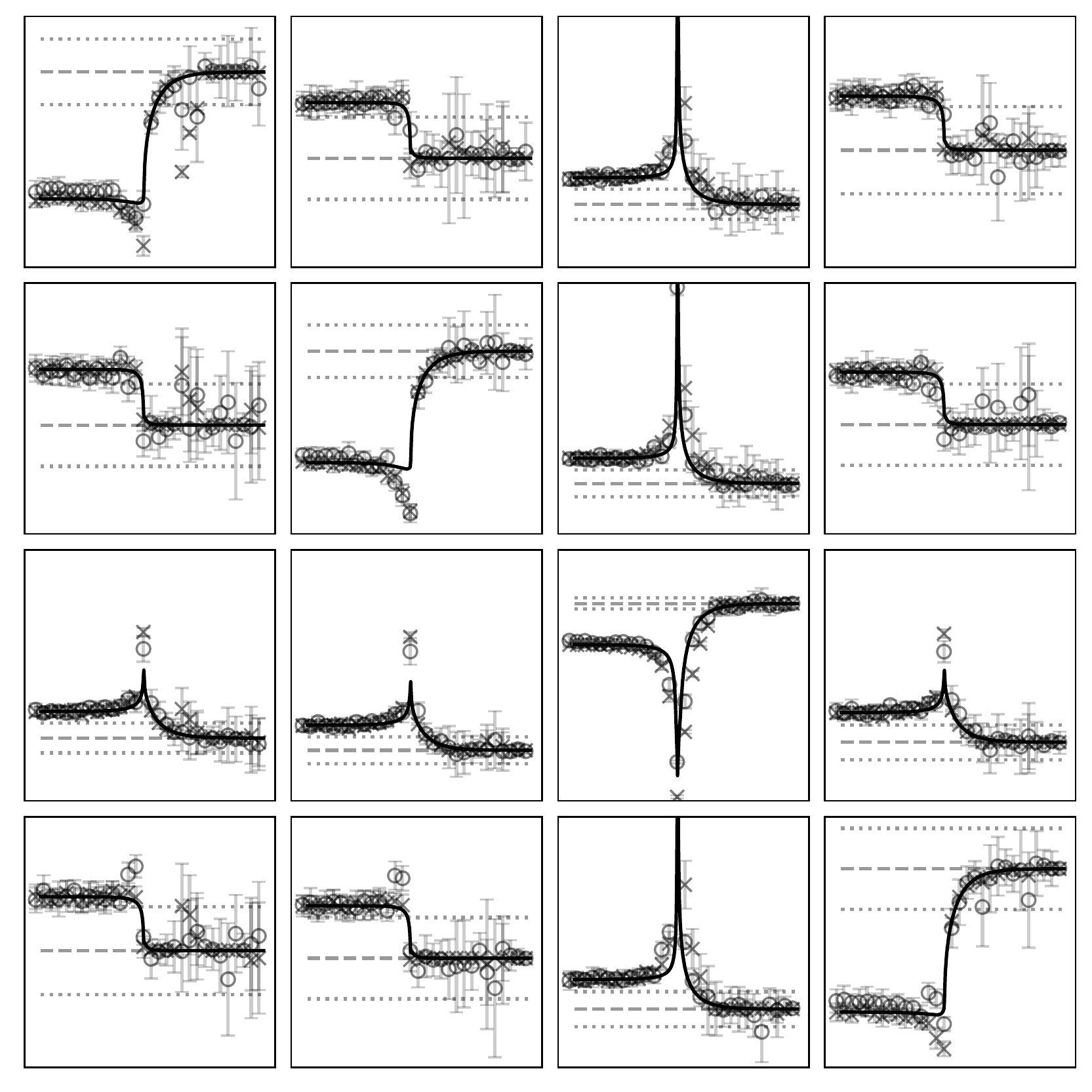}
    \caption{Evolution of the individual Slutsky matrix entries with $c$ for $M=4$ product, $k=2$, $w=10$, $\mathbf{p} = [2.2,2.1,1.6,2.3]$, $a_i = 1$ from theory at $\beta \to \infty$ (continuous lines) and numerical experiments for $\beta = 4$, $N = 16$, calculated using both the pathwise derivative estimates introduced Appendix \ref{appendix:mc} (circles) and the fluctuation-response relations (crosses), averaged over all agents, errorbars indicating one standard deviation. The dashed horizontal line indicates 0, while dotted lines correspond to $\pm 0.1$ as no vertical scale is shown for visual clarity. Opposing entries have identical vertical scales as to highlight the strong asymmetry of some entries (e.g. 13 and 31) close to the herding transition.}
    \label{fig:S_full}
\end{figure}

Using a simple interacting model, we have shown that introducing herding in the problem leads to a concentration transition and radical changes in the way agents allocate their budget among the $M$ available goods. We now set out to evaluate the impact of such a transition on the Slutsky matrix and its properties, in particular its negative semi-definiteness and symmetry.

Whereas the grand-canonical theory allowed us to calculate $\langle x_i^\alpha \rangle = \overline{x}_i^*$ for the entire range of $\beta$ and $c$ self-consistently, the absence of an explicit expressions prevents us from directly computing $S_{ij}^\alpha$. Instead, because we have found that results are largely independent of $\beta$ provided $c$ is rescaled as $c/c_\mathrm{crit}$, we may gain insight from the Gaussian approximation of the Slutsky matrix introduced in Section \ref{sec:many_agents} and valid for $\beta \to \infty$. Equation~\eqref{equ:Slutsky_T=0} can now be made explicit and writes (see Appendix \ref{appendix:SlutskyZeroT})
\begin{equation}
\begin{aligned}
        S_{ij}^\alpha = - \beta \sum_\gamma \bigg(& \frac{a_\ell}{p_\ell \overline{x}_\ell^*} [ 1 + c (\overline{x}_\ell^*)^k(1+k\log \overline{x}_\ell^*)] \\ 
        &+ (1- \delta_{\alpha \gamma}) \frac{a_j}{p_j \overline{x}_j^*}[1+c(\overline{x}_j^*)^k] \bigg) {C}_{ij}^{\alpha \gamma},
\end{aligned}
\end{equation}
for any $\ell = 1\dots M$. As the most probable values $\overline{\mathbf{x}}^*$ can be obtained by solving Eq.~\eqref{equ:x_i_betainf}, the last step is to invert the Hessian of the utility function in order to compute the covariance matrix $\mathbf{C}$. Due to the homogeneous nature of the interactions, the Hessian has a very regular structure and may thus be inverted explicitly, as detailed in Appendix \ref{appendix:SlutskyZeroT}. Doing so and replacing in Eq.~\eqref{equ:correls}, we find that $\mathbf{C}$ has the following form 
\begin{equation*}
    C_{ij}^{\alpha \gamma} = \frac{1}{\beta}\bigg[\varphi_{ij} \delta_{\alpha \gamma} + \frac{1}{N}\psi_{ij} \bigg],
\end{equation*}
where $\varphi_{ij}$ and $\psi_{ij}$ are both symmetric and $\mathcal{O}(1)$ in $N$, and, of course, depend on $\overline{\mathbf{x}}^*$. Such correlations, which can be checked to match numerical simulations very well, finally yield an identical Slutsky matrix for all agents,
\begin{equation}
\begin{aligned}
     S_{ij} = - \frac{a_j}{p_j \overline{x}_j^*} \big[ &kc(\varphi_{ij} + \psi_{ij}) (\overline{x}_j^*)^{k} \log \overline{x}_j^*\\
    & + [1+c(\overline{x}_j^*)^k] \varphi_{ij}\big].
\end{aligned}
\end{equation}
Together with Eq.~\eqref{equ:Slutsky_noderivatives}, this is a key result of our work. 

\begin{figure}
    \centering
    \includegraphics[width=\linewidth]{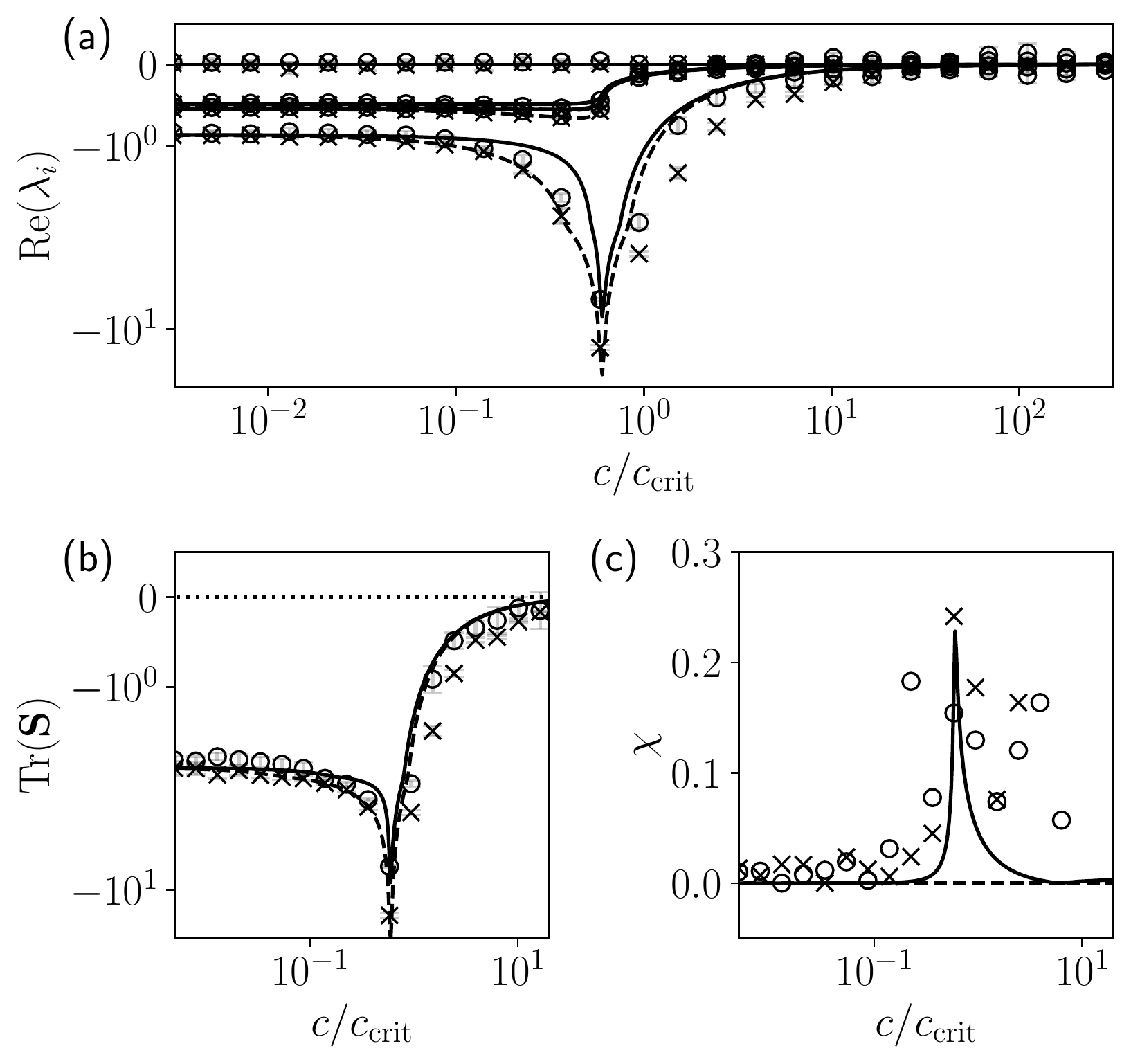}
    \caption{Properties of the individual Slutsky matrix shown in Fig.~\ref{fig:S_full}.  (a) Real part of the Slutsky matrix eigenvalues $\lambda_i$. (b) Trace of the matrix, dotted line indicating zero. (c) Asymmetry measure $\chi$. The analytical results for the aggregate matrix $\bm{\mathcal{S}}$ are given by the dashed lines for comparison (note in particular that $\bm{\mathcal{S}}$ remains symmetric for all $c$).}
    \label{fig:S_properties}
\end{figure}

Thus far, we have considered the case where all agents and goods are  identical. In the present context, however, it is more illustrative to break the symmetry between products by introducing heterogeneous $p_i$ and $a_i$, leading to clear preferences between products. An example with $M = 4$ goods is given in Fig.~\ref{fig:S_full}, showing a very good agreement between the fully rational theory and numerical simulations with finite $\beta$ and $N$ sufficiently far from the transition region.\footnote{Close to the transition, agents flip-flop between different basket compositions (Fig.~\ref{fig:Transition_betafinite}(b)), which leads to anomalously strong fluctuations and corrections to the Laplace saddle point method used in all our analytical calculations. This is a well known effect in statistical physics, which leads to interesting phenomena in their own right, but that we do not explore further here.} Upon inspection of opposing entries, it quickly appears that the symmetry of the matrix is not satisfied near the critical value of $c$ where condensation first occurs. Note that all matrix entries quickly become very small once the system has entered the concentrated phase. This can be understood for large $c$ by making the \textit{ansatz} $\overline{x}_i^* = w/p_i - \frac{1}{c}\sum_{j\neq i} y_j$ for the dominant product and $\overline{x}_j^* = y_j/c$ for others. Plugging such a guess  in Eq.~\ref{equ:x_i_betainf} indeed solves the equations, and predicts $S_{ij} \sim 1/c$ to leading order in $c^{-1}$. 

The first property of the matrix that interests us is its spectrum, and in particular the non-positivity of its eigenvalues. As shown in Fig~\ref{fig:S_properties}(a), the fully rational theory provides a very good description of the matrix eigenvalues, which remain non-positive for the entire range of $c$. The leading eigenvalue actually peaks close to the transition. Consistent with the decay of the matrix entries themselves, the magnitude of the eigenvalues also vanish as $c$ increases beyond the transition. As a result, our theory and numerical experiments show that the Slutsky matrix does remain negative semi-definite for all value of $c$ and $\beta$. The main consequence of the herding transition on the spectrum is the decay in the magnitude of eigenvalues, as confirmed by looking at the trace of the matrix shown in Fig.~\ref{fig:S_properties}(b).

The other essential property of the Slutsky matrix  is its symmetry. Due to the decay in the magnitude of the matrix entries once the system has entered the herding phase, this property becomes difficult to measure as $c$ is increased beyond $c_\mathrm{crit}$. Indeed, as both $S_{ij}$ and $S_{ji}$ become very small, any finite numerical error $\varepsilon$ affecting either entries will result in a very large relative asymmetry, at which point most common measures of asymmetry will fail. To minimize the impact of such numerical errors on the conclusions of our study, we propose an asymmetry measure $\chi$ defined as
\begin{equation}
    \chi = \left| \frac{\sum_{\alpha} \sum_{j<i} (S_{ij}^\alpha - S_{ji}^\alpha)}{\sum_{\alpha} \sum_{j<i} (S_{ij}^\alpha + S_{ji}^\alpha)} \right|,
\end{equation}
which should be equal to zero for symmetric matrices, and diverge in the anti-symmetric case. Employing this metric with the $\beta \to \infty$ theoretical Slutsky matrix, we find that the matrix is very close to (but not exactly) symmetric far from the condensation transition. In the vicinity of $c_\mathrm{crit}$ however, strong interactions give rise to a significant value of $\chi$, contradicting the conventional lore even in the fully rational case. This theoretical result, compared with numerical results, is shown in Fig.~\ref{fig:S_properties}(c), while also visible in  Fig.~\ref{fig:S_full}. 

Note that, as expected given our choice of identical agents with identical budgets, the previously introduced \textit{aggregate} Slutsky matrix $\bm{\mathcal{S}}$ remains symmetric even as the concentration of choice occurs. The theoretical results for this alternate definition are shown by the dashed lines in Fig.~\ref{fig:S_properties}, where it is clear that the asymmetry measure $\chi$ is always zero. Interestingly, the eigenvalues of the matrix are very similar for the individual ($S_{ij}$) and aggregate ($\mathcal{S}_{ij}$) definitions.

Regardless of the metric, Fig.~\ref{fig:S_properties} also illustrates the discrepancy between the equilibrium theory we have devised and the numerical measurements in the transition region. As expected, the system indeed takes a very long time to reach the Boltzmann-Gibbs distribution when the transition occurs. These non-equilibrium effects are also reflected in the difference between the numerical measurements obtained with finite differences and those calculated using the fluctuation-response relations and the associated ``thermodynamic'' expression of the Slutsky matrices. We expect that such effects will also be present in real empirical data, specially if herding effects bring the system close to a transition point, as seemed to be the case in the Salganik \textit{et al.} experiment \cite{salganik2006experimental, Borghesi2007}.

\section{Discussion} 
\label{sec:Discussion}

Before concluding our study, let us examine two subtle points that the above analysis has treated in a somewhat cavalier way.

\subsection{Global vs. Individual Utilities}
\label{sec:global_vs_individual}

From the very start of our statistical mechanics description of the multi-agent problem, we have emphasized that we are assuming that individual agents change their basket of goods according to the change of the global utility of the population rather than of their own utility. In other words, agents also take into account the change of utility of others when they update their choices. The main motivation behind this specification is that the dynamics will then spontaneously reach a Boltzmann-Gibbs equilibrium measure. Indeed, following an agent-based framework where individuals set out to improve their own utility function, including the herding component $c (\overline x)^k$ would bring us in the realm of non-Hamiltonian dynamics, for which general analytical tools are still lacking. Natural questions are now (i) how may the current model still be interpreted at the agent level and (ii) how is the phenomenology of the system impacted if we abandon the Hamiltonian description and take agents to maximize their own utilities.

Regarding the former question, we start by writing the change in global utility $\Delta U$ if the randomly selected agent $\gamma$ in the Monte Carlo dynamics accepts the proposed basket of goods $\mathbf{x}^\gamma + \Delta \mathbf{x}$. Assuming $N \gg 1$, we have
\begin{equation}
    \Delta U \approx \sum_{i=1}^M \frac{a_i}{x_i^\gamma} \Big(1+c[1+ k\overline{\log x_i}](\overline{x}_i)^k\Big) \Delta x_i,
\end{equation}
where we remind the reader that the overline notation refers to arithmetic averages over the agents. This equation shows that the dynamics can in fact be interpreted in terms of agents only concerned with their own utility, albeit with modified values of the interaction parameter $c$ ($c \to c_i= c [1+k\overline{\log x_i}]$). The individual utility thus becomes configuration dependent, but only through aggregate quantities (the average of the logarithm of $x_i$). This logarithm correction does lead to a supplementary penalty for products that have lost the favor of the crowd, but is not expected to radically change the concentration phenomenon reported in the above sections. 

Consistent with this interpretation, simulating the system with a decision rule based on the individual utility but with a constant $c$ -- i.e. placing ourselves in a non-Hamiltonian setting -- indeed results in a largely unchanged phenomenology, with the same herding transition as was observed in the Hamiltonian case. As expected, the absence of this logarithmic penalty for small $\overline{x}_i$ pushes the transition to slightly higher values of $c$ and leads to more volatile Monte Carlo trajectories, with the appearance of more frequent switches in the vicinity of the transition. Although the absence of a solid theoretical framework to describe the steady state in that case prevents us from drawing definitive conclusions at this stage, we conjecture that most of the results obtained at equilibrium regarding the Slutsky matrix continue to hold, with only minor quantitative modifications.

Alternatively, one could also purposefully write a global utility for which the detailed-balance condition matches the maximization of an agent-specific utility. To do so, the interaction term must be entirely symmetric in the sense that the change in global utility is identical regardless of the randomly selected agent (which was not the case with the previously studied model). For instance, one could take
\begin{equation}
    U(\{\mathbf{x}^\alpha\}) = \sum_{i,\alpha} a_i^\alpha \log x_i^\alpha + \sum_{\substack{i,\alpha,\gamma \\ \gamma \neq \alpha}} J_i^{\alpha \gamma} (x_i^\alpha)^\rho (x_i^\gamma)^\rho,
    \label{equ:U_new}
\end{equation}
with $\mathbf{J}_i$ a \textit{symmetric} interaction matrix. The mean-field approximation of this model $J_i^{\alpha \gamma} = a_i J/N$ is studied in the limit $\beta \to \infty$ in Appendix~\ref{appendix:Hamiltonian_U}. For $\rho > \frac{1}{2}$, concentration will occur for sufficiently large values of $J$, and we therefore expect our results for the Slutsky matrix properties to remain largely unchanged.

In the case where $\mathbf{J}_i$ is not symmetric, the problem may no longer be treated with the standard techniques of equilibrium statistical mechanics. The system can then still be simulated with a decision rule maximizing the individual utility of the agent, but we can no longer assume that the system reaches an equilibrium given by the Boltzmann-Gibbs distribution.

\begin{figure}
    \centering
    \includegraphics[width=\linewidth]{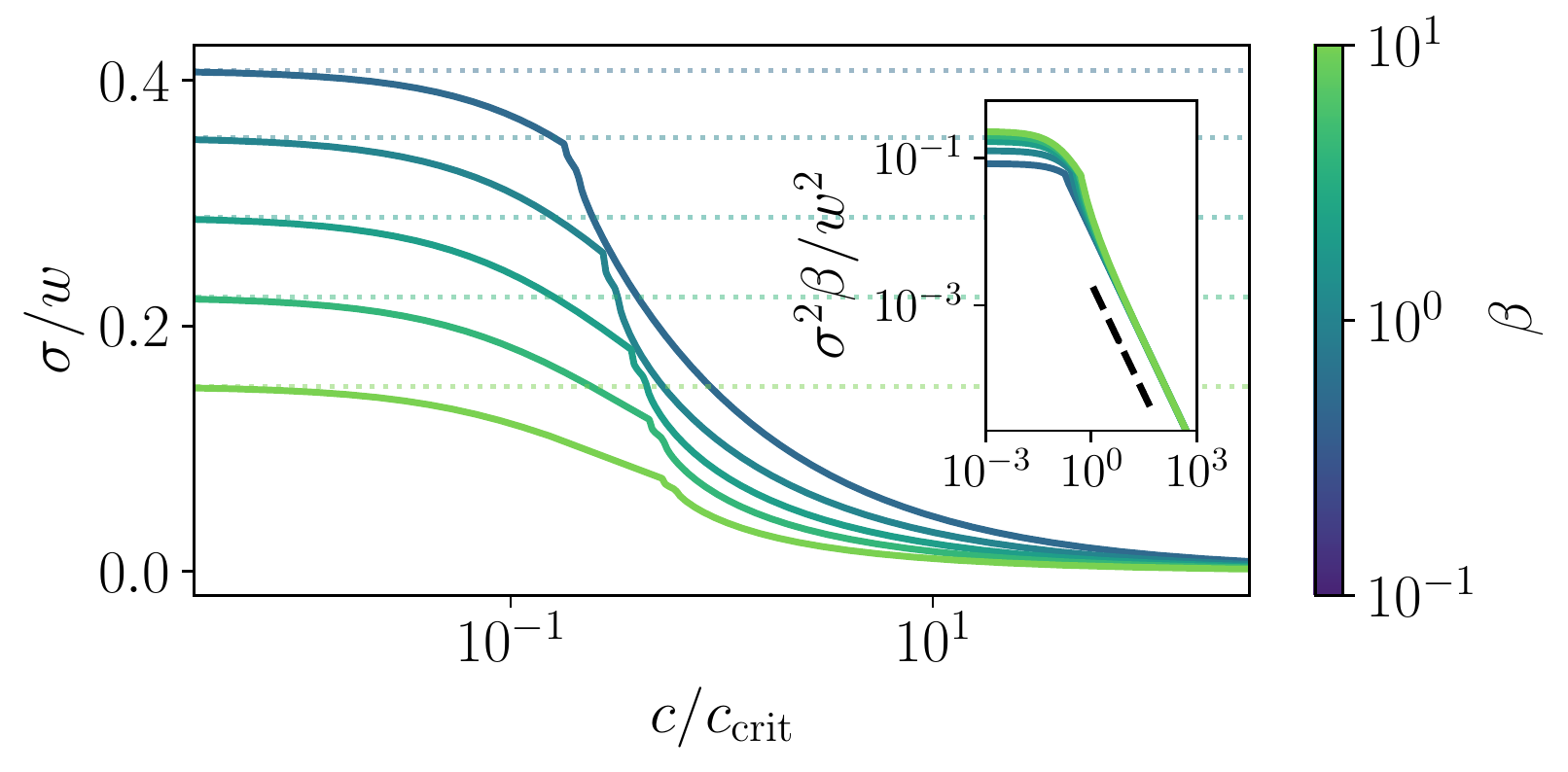}
    \caption{Evolution of the relative fluctuations in the realized budget in the grand-canonical ensemble as a function of $c/c_\mathrm{crit}$ for $M=4$, $k=2$, $w=10$, $a_i=p_i=1$ and $\beta = 0.5,1,2,4,10$, dotted lines corresponding to the unconcentrated solution. Inset: log-log scale showing $\sigma^2 \beta$ decreasing as $1/c$ (dashed line) for $c \gtrsim c_\mathrm{crit}$.}
    \label{fig:Equiv}
\end{figure}

\subsection{Equivalence of Ensembles}
\label{sec:equivalence}
In order to study analytically the equilibrium properties of our interacting set of agents and to determine when the herding transition occurs, we had to relax the budget constraint and place ourselves in the ``grand-canonical'' ensemble. As previously mentioned, the equivalence of results between the canonical and grand-canonical ensembles is not guaranteed \textit{a priori}, although in the present case the analytical (grand-canonical) results appear to match (canonical) simulations extremely well. 

To formally assess the possible differences between the two ensembles, the budget fluctuations in the grand-canonical ensemble must be studied. The ensemble equivalence corresponds to cases where the variance of the realized budget are vanishingly small. Hence, we set out to compute
\begin{equation}
    \sigma^2 = \bigg \langle \Big( \sum_{i=1}^M x_i^\alpha p_i - w \Big)^2 \bigg \rangle
\end{equation}
in the grand-canonical ensemble. This computation may be performed by introducing  small heterogeneous perturbations to the chemical potential, $\mu \to \mu + \delta \mu^\alpha$ and differentiating $\log \mathcal{Z}_N$ with respect to $\delta \mu^\alpha$. The calculation, detailed in Appendix \ref{appendix:ensembles}, finally yields
\begin{equation}
    \sigma^2 = \sum_{i=1}^M \frac{(\overline{x}_i^* p_i)^2}{1+\beta a_i[1+c(\overline{x}_i^*)^k]}.
\end{equation}
The evolution of this quantity for different values of $\beta$ as a function of $c$ in the previously discussed case $a_i = p_i = 1$ is shown in Fig.~\ref{fig:Equiv}. As expected, this quantity vanishes for $\beta \to \infty$, as well as for $c\to \infty$. In the $c < c_\mathrm{crit}$ region, where fluctuations are expected to be the strongest and thus where the difference between the ensembles is expected to be the largest, the non-condensed solution (dotted lines on Fig.~\ref{fig:Equiv}) gives $\sigma^2 \sim \mathcal{O}(M^{-1})$. As such, the grand-canonical and canonical ensembles will become strictly equivalent only in the limit $M\to \infty$. This being said, the Hessian of the free energy being independent of the chemical potential suggests that despite this apparent absence of strict equivalence between the ensembles for finite $M$, the onset of the transition appears to be largely unaffected. In any case, although precise measures of the fluctuations of the system in the grand-canonical ensemble should be inaccurate for smaller values of $\beta$, this difference will not change the key results presented in this work. Besides, one should expect that some amount of budget fluctuations are indeed present in the real world!

\section{Conclusion}
\label{sec:Conclusion}

Let us summarize what we have achieved in this paper. By introducing a rationality parameter (or ``intensity of choice'') $\beta$ to account for the fact that agents are not strict utility maximizers, we have first reformulated the Slutsky equation within a general ``fluctuation-response'' framework, which allows one to express the Slutsky matrix in terms of consumption fluctuations only, without having to measure changes of consumptions when prices are slightly modified. 

We have then shown that irrationality does {\it not} necessarily result in the breakdown of the symmetry of the Slutsky matrix. As a result, the hypothetical symmetry of empirically measured Slutsky matrices {\it cannot} be used as a general argument against bounded rationality ~\cite{ekeland2011}. 

When accounting for herding within large assemblies of agents, we found that symmetry is no longer guaranteed in general. Introducing a simple model of utility with interactions, we have indeed shown using the powerful methods offered by statistical mechanics that a concentration transition may occur, at which point strong selection of goods occurs. At the transition, the individual Slutsky matrix becomes markedly asymmetric, although the Slutsky matrix constructed using aggregate consumption can still remain symmetric when all agents are identical. Hence our result is not necessarily incompatible with existing work on non-unitary households \cite{Browning1998}. From simulations, we also found that asymmetry is further amplified by out-of-equilibrium effects near the critical point. In line with standard consumer choice theory and most empirical studies, our model preserves the negative semi-definiteness of the Slutsky matrix regardless of interactions. Nonetheless and although not studied here, it should be noted that introducing some interactions in between products (representing redundancy or complementarity for example) in a similar framework appears to lead to positive eigenvalues \cite{lecointre2019texture}. 

Central to our theoretical framework, the assumption that the system may be written in Hamiltonian form, meaning agents are not strictly individualistic when interactions are included, does not appear to be necessary for our results to hold. This being said, further investigations in particular regarding imitation-lead switches in an out-of-equilibrium model would be a natural path to explore, although we expect doing so may only lead to stronger asymmetry of the Slutsky matrix. Considering different forms of utility and their impact on the symmetry of the Slutsky matrix is another potential aspect to expand upon. In particular, adding disorder in the interactions (for example by taking $\mathbf{J}_i$ to be a random matrix in Eq.~\eqref{equ:U_new}) would likely lead to very interesting effects. 

Of course, further empirical studies on the properties of the Slutsky matrix would be of great interest, in particular to contrast our model of interacting, bounded rational agents with the recent sparsity-based approach of Gabaix \cite{gabaix2014sparsity}. To this end, we believe that fluctuation-response relations such as those presented in this work could be a very valuable econometric tool. Given the difficulty of conducting repeatable experiments in socioeconomic systems, the ability to estimate derivative quantities {\it without} actually requiring prices to change appears quite promising.

\section*{Acknowledgments}
We warmly thank P. Lecointre and P.-P. Crépin who participated in the early stages of this project. It is also our pleasure to thank C. Aubrun, G. Biroli, T. Dessertaine, S. Lakhal and P. Mergny for fruitful discussions. This research was conducted within the Econophysics \& Complex Systems Research Chair, under the aegis of the Fondation du Risque, the Fondation de l’Ecole polytechnique, the Ecole polytechnique and Capital Fund Management. The authors have no conflicts to disclose.

\bibliographystyle{unsrt}
\bibliography{biblio}

\begin{thebibliography}{10}

\bibitem{simon1972theories}
Herbert~A Simon.
\newblock Theories of bounded rationality.
\newblock {\em Decision and organization}, 1(1):161--176, 1972.

\bibitem{arthur1994inductive}
W.~Brian Arthur.
\newblock Inductive reasoning and bounded rationality.
\newblock {\em The American Economic Review}, 84(2):406--411, 1994.

\bibitem{gigerenzer2002bounded}
Gerd Gigerenzer and Reinhard Selten.
\newblock {\em Bounded rationality: The adaptive toolbox}.
\newblock MIT press, 2002.

\bibitem{brockdurlauf2001}
William~A. Brock and Steven~N. Durlauf.
\newblock Discrete choice with social interactions.
\newblock {\em The Review of Economic Studies}, 68(2):235--260, 2001.

\bibitem{Kirman2010}
A.P. Kirman.
\newblock {\em Complex Economics: Individual and Collective Rationality.}
\newblock Routledge, 2010.

\bibitem{Bouchaud2013}
J.-P. Bouchaud.
\newblock Crises and collective socio-economic phenomena: simple models and
  challenges.
\newblock {\em Journal of Statistical Physics}, 151(3-4):567–606, 2013.

\bibitem{morelli2020confidence}
Federico~Guglielmo Morelli, Michael Benzaquen, Marco Tarzia, and Jean-Philippe
  Bouchaud.
\newblock Confidence collapse in a multihousehold, self-reflexive dsge model.
\newblock {\em Proceedings of the National Academy of Sciences},
  117(17):9244--9249, 2020.

\bibitem{bouchaud2001power}
Jean-Philippe Bouchaud.
\newblock Power laws in economics and finance: some ideas from physics.
\newblock {\em Quantitative finance}, 1(1):105, 2001.

\bibitem{axtell2001zipf}
Robert~L Axtell.
\newblock Zipf distribution of us firm sizes.
\newblock {\em Science}, 293(5536):1818--1820, 2001.

\bibitem{farmer2008power}
J~Doyne Farmer and John Geanakoplos.
\newblock Power laws in economics and elsewhere.
\newblock In {\em Santa Fe Institute}, 2008.

\bibitem{gabaix2009power}
Xavier Gabaix.
\newblock Power laws in economics and finance.
\newblock {\em Annu. Rev. Econ.}, 1(1):255--294, 2009.

\bibitem{gabaix2011granularity}
Xavier Gabaix.
\newblock The granular origins of aggregate fluctuations.
\newblock {\em Econometrica}, 79(3):733--772, 2011.

\bibitem{kaplan2018monetary}
Greg Kaplan, Benjamin Moll, and Giovanni~L Violante.
\newblock Monetary policy according to {HANK}.
\newblock {\em American Economic Review}, 108(3):697--743, 2018.

\bibitem{abel1990asset}
Andrew Abel.
\newblock Asset prices under habit formation and catching up with the
  {Joneses}.
\newblock Technical report, March 1990.

\bibitem{campbell1999habit}
John~Y. Campbell and John~H. Cochrane.
\newblock By force of habit: A consumption‐based explanation of aggregate
  stock market behavior.
\newblock {\em Journal of Political Economy}, 107(2):205--251, 1999.

\bibitem{carroll2000saving}
Christopher~D. Carroll, Jody Overland, and David~N. Weil.
\newblock Saving and growth with habit formation.
\newblock {\em American Economic Review}, 90(3):341--355, June 2000.

\bibitem{moran2020by}
Jos{\'e} Moran, Antoine Fosset, Davide Luzzati, Jean-Philippe Bouchaud, and
  Michael Benzaquen.
\newblock By force of habit: Self-trapping in a dynamical utility landscape.
\newblock {\em Chaos: An Interdisciplinary Journal of Nonlinear Science},
  30(5):053123, 2020.

\bibitem{slutsky1915}
Eugenio Slutsky.
\newblock Sulla teoria del bilancio del consumatore.
\newblock {\em Giornale degli economisti e rivista di statistica}, 1915.

\bibitem{Brown1972}
Alan Brown and Angus Deaton.
\newblock Surveys in applied economics: models of consumer behaviour.
\newblock {\em The Economic Journal}, 82(328):1145--1236, 1972.

\bibitem{Samuelson1974}
Paul~A Samuelson.
\newblock Complementarity: An essay on the 40th anniversary of the
  {Hicks-Allen} revolution in demand theory.
\newblock {\em Journal of Economic literature}, 12(4):1255--1289, 1974.

\bibitem{Barten1982}
Anton~P Barten and Volker B{\"o}hm.
\newblock Consumer theory.
\newblock {\em Handbook of mathematical economics}, 2:381--429, 1982.

\bibitem{Grandville1989}
Olivier de~La~Grandville.
\newblock In quest of the {Slutsky} diamond.
\newblock {\em The American Economic Review}, pages 468--481, 1989.

\bibitem{fuleihan1968empirical}
Joseph~S Fuleihan.
\newblock {\em The empirical estimation of substitution terms from demand
  analysis}.
\newblock PhD thesis, Iowa State University, 1967.

\bibitem{Barten1967}
Anton~P Barten.
\newblock Evidence on the {Slutsky} conditions for demand equations.
\newblock {\em The Review of Economic and Statistics}, pages 77--84, 1967.

\bibitem{Berndt1977}
Ernst~R Berndt, Masako~N Darrough, and W~Erwin Diewert.
\newblock Flexible functional forms and expenditure distributions: An
  application to canadian consumer demand functions.
\newblock {\em International Economic Review}, pages 651--675, 1977.

\bibitem{blundell1993we}
Richard Blundell, Panos Pashardes, and Guglielmo Weber.
\newblock What do we learn about consumer demand patterns from micro data?
\newblock {\em The American Economic Review}, pages 570--597, 1993.

\bibitem{Browning1998}
M.~Browning and P.-A. Chiappori.
\newblock Efficient intra-household allocations: A general characterization and
  empirical tests.
\newblock {\em Econometrica}, page 1241–1278, 1998.

\bibitem{Mirowski1998}
Philip Mirowski and D~Wade Hands.
\newblock A paradox of budgets: the postwar stabilization of american
  neoclassical demand theory.
\newblock {\em History of Political Economy}, page 260, 1998.

\bibitem{Diewert1977}
W~Erwin Diewert.
\newblock Generalized slutsky conditions for aggregate consumer demand
  functions.
\newblock {\em Journal of Economic Theory}, 15(2):353--362, 1977.

\bibitem{lechene2011noncooperative}
Val{\'e}rie Lechene and Ian Preston.
\newblock Noncooperative household demand.
\newblock {\em Journal of Economic Theory}, 146(2):504--527, 2011.

\bibitem{gabaix2014sparsity}
Xavier Gabaix.
\newblock A sparsity-based model of bounded rationality.
\newblock {\em The Quarterly Journal of Economics}, 129(4):1661--1710, 2014.

\bibitem{aguiar2017slutsky}
Victor~H Aguiar and Roberto Serrano.
\newblock Slutsky matrix norms: The size, classification, and comparative
  statics of bounded rationality.
\newblock {\em Journal of Economic Theory}, 172:163--201, 2017.

\bibitem{aguiar2018classifying}
Victor~H Aguiar and Roberto Serrano.
\newblock Classifying bounded rationality in limited data sets: a {Slutsky}
  matrix approach.
\newblock {\em SERIEs}, 9(4):389--421, 2018.

\bibitem{Mas1995}
Andreu Mas-Colell, Michael~Dennis Whinston, Jerry~R Green, et~al.
\newblock {\em Microeconomic theory}, volume~1.
\newblock Oxford university press New York, 1995.

\bibitem{anderson1992discrete}
Simon~P. Anderson, Andre de~Palma, and Jacques-Francois Thisse.
\newblock {\em Discrete Choice Theory of Product Differentiation}.
\newblock The MIT Press, 1992.

\bibitem{ekeland2011}
Jon Elster and Ivar Ekeland.
\newblock {\em Th{\'e}orie {\'e}conomique et rationalit{\'e}}.
\newblock Vuibert, 2011.

\bibitem{marconi2008fluctuation}
Umberto Marini~Bettolo Marconi, Andrea Puglisi, Lamberto Rondoni, and Angelo
  Vulpiani.
\newblock Fluctuation--dissipation: response theory in statistical physics.
\newblock {\em Physics reports}, 461(4-6):111--195, 2008.

\bibitem{veblen}
Thorstein Veblen.
\newblock {\em The theory of the leisure class}.
\newblock 1899.

\bibitem{bass1969new}
Frank~M Bass.
\newblock A new product growth for model consumer durables.
\newblock {\em Management science}, 15(5):215--227, 1969.

\bibitem{salganik2006experimental}
Matthew~J Salganik, Peter~Sheridan Dodds, and Duncan~J Watts.
\newblock Experimental study of inequality and unpredictability in an
  artificial cultural market.
\newblock {\em Science}, 311(5762):854--856, 2006.

\bibitem{gordon2009discrete}
Mirta~B Gordon, Jean-Pierre Nadal, Denis Phan, and Viktoriya Semeshenko.
\newblock Discrete choices under social influence: Generic properties.
\newblock {\em Mathematical Models and Methods in Applied Sciences},
  19(supp01):1441--1481, 2009.

\bibitem{stracca2005keeping}
Ali al~Nowaihi and Livio Stracca.
\newblock {Keeping up with the Joneses, reference dependence, and equilibrium
  indeterminacy}.
\newblock Working Paper Series 444, European Central Bank, February 2005.

\bibitem{Borghesi2007}
Christian Borghesi and Jean-Philippe Bouchaud.
\newblock Of songs and men: a model for multiple choice with herding.
\newblock {\em Quality {\&} Quantity}, 41(4):557--568, March 2007.

\bibitem{lucas2022non}
Andrew Lucas.
\newblock Non-equilibrium phase transitions in competitive markets caused by
  network effects.
\newblock {\em arXiv preprint arXiv:2204.05314}, 2022.

\bibitem{lecointre2019texture}
Pierre Lecointre.
\newblock {\em Texture-induced hydrophobicity, and some collective effects in
  granular matter and economics}.
\newblock PhD thesis, Institut Polytechnique de Paris, 2019.

\bibitem{newman1999monte}
Mark~EJ Newman and Gerard~T Barkema.
\newblock {\em Monte Carlo methods in statistical physics}.
\newblock Clarendon Press, 1999.

\bibitem{glasserman1991gradient}
Paul Glasserman and Yu-Chi Ho.
\newblock {\em Gradient estimation via perturbation analysis}, volume 116.
\newblock Springer Science \& Business Media, 1991.

\bibitem{glasserman2013monte}
Paul Glasserman.
\newblock {\em Monte Carlo methods in financial engineering}, volume~53.
\newblock Springer Science \& Business Media, 2013.

\bibitem{grauwin2009competition}
S{\'e}bastian Grauwin, Eric Bertin, R{\'e}mi Lemoy, and Pablo Jensen.
\newblock Competition between collective and individual dynamics.
\newblock {\em Proceedings of the National Academy of Sciences},
  106(49):20622--20626, 2009.

\end{thebibliography}

\clearpage

\onecolumngrid

\small

\appendix

\section{General ``Thermodynamic'' Relations}
\subsection{Single Agent Slutksy Matrix}
\label{appendix:SingleAgentSlutsky}

Assuming the equilibrium distribution is reached, Eq.~\eqref{equ:Slutsky_canonical} can be rearranged to be rewritten through correlation functions. Indeed, from the quotient rule the first term becomes
\begin{equation}
    \frac{\partial}{\partial p_j} \langle x_i \rangle = \frac{1}{Z} \int \dd \mathbf{x} \, x_i x_j \e^{\beta u(\mathbf{x})} \delta'(\mathbf{x}\cdot \mathbf{p} - w) - \frac{\langle x_i \rangle}{Z} \int \dd \mathbf{x} \, x_j \e^{\beta u(\mathbf{x})} \delta'(\mathbf{x}\cdot \mathbf{p} - w).
\end{equation}
Now, using the fact that
\begin{equation}
    \delta'(\mathbf{x}\cdot \mathbf{p} - w) = -\partial_w \delta(\mathbf{x}\cdot \mathbf{p} - w),
\end{equation}
then for a test function $O(\mathbf{x})$
\begin{equation}
    \frac{1}{Z} \int \dd \mathbf{x} \, O(\mathbf{x}) \, \e^{\beta u(\mathbf{x})} \delta'(\mathbf{x}\cdot \mathbf{p} - w) = - \frac{\partial}{\partial w} \langle O(\mathbf{x}) \rangle - \langle O(\mathbf{x}) \rangle \partial_w \log Z.
\end{equation}
Therefore bringing everything together we have
\begin{equation}
    S_{ij} = \langle x_i \rangle \partial_w \langle x_j \rangle + \langle x_i \rangle \langle x_j \rangle \partial_w \log Z - \partial_w \langle x_i x_j \rangle - \langle x_i x_j \rangle \partial_w \log Z + \langle x_j \rangle \partial_w \langle x_i \rangle,
\end{equation}
or in a more compact form, with $\langle x_i x_j \rangle_c = \langle x_i x_j \rangle - \langle x_i \rangle \langle x_j \rangle$ and $\Gamma = \partial_w \log Z$,
\begin{equation}
    S_{ij} = -\Gamma \langle x_i x_j \rangle_c - \partial_w \langle x_i x_j \rangle_c.
\end{equation}

\subsection{Many Agent Slutsky Matrix}
\label{appendix:MultiAgentSlutsky}
The agent-specific Slutsky matrix is defined as
\begin{equation}
    S_{ij}^\alpha = \frac{\partial}{\partial p_j} \langle x_i^\alpha \rangle + \langle x_j^\alpha \rangle \frac{\partial}{\partial w^\alpha} \langle x_i^\alpha \rangle.
\end{equation}
Once again assuming equilibrium is reached, we wish to rewrite this expression through correlations to retrieve some information on the symmetry of the matrix. As before, we look at the first term in the above equation,
\begin{equation}
    \frac{\partial}{\partial p_j} \langle x_i^\alpha \rangle = \frac{1}{Z_N} \int \mathcal{D} \mathbf{x} \, x_i^\alpha \e^{\beta U(\{\mathbf{x}^\gamma\})} \frac{\partial}{\partial p_j} \prod_\gamma \delta(\mathbf{x}^\gamma \cdot \mathbf{p} - w^\gamma) - \frac{\langle x_i^\alpha \rangle}{Z_N} \int \mathcal{D} \mathbf{x} \, \e^{\beta U(\{\mathbf{x}\}^\gamma)} \frac{\partial}{\partial p_j} \prod_\gamma \delta(\mathbf{x}^\gamma \cdot \mathbf{p} - w^\gamma),
    \label{equ:x_i_alpha_p_j}
\end{equation}
with hereafter the shorthand notation $\int \mathcal{D}\mathbf{x} := \int_+ \prod_{\alpha,i} \dd x_i^\alpha$. Now, by the generalized chain rule we have
\begin{equation}
    \frac{\partial}{\partial p_j} \prod_\gamma \delta(\mathbf{x}^\gamma \cdot \mathbf{p} - w^\gamma) = -\sum_\gamma \frac{\partial}{\partial w^\gamma} x_j^\gamma \prod_\eta \delta(\mathbf{x}^\eta \cdot \mathbf{p} - w^\eta),
\end{equation}
which may be plugged back into Eq.~\eqref{equ:x_i_alpha_p_j}. Noticing that 
\begin{equation}
    \Big\langle \frac{\partial}{\partial w^\gamma} O(\mathbf{x^\eta}) \Big \rangle = \frac{\partial}{\partial w^\gamma} \langle O(\mathbf{x}^\eta) \rangle + \Gamma_\gamma \langle O(\mathbf{x}^\eta) \rangle,
\end{equation}
with $\Gamma_\gamma = \frac{\partial}{\partial w^\gamma} \log Z_N$, then
\begin{equation}
    \frac{\partial}{\partial p_j} \langle x_i^\alpha \rangle = - \sum_\gamma \left[ \Gamma_\gamma \langle x_i^\alpha x_j^\gamma\rangle_c + \frac{\partial}{\partial w^\gamma} \langle x_i^\alpha x_j^\gamma \rangle -\langle x_i^\alpha \rangle  \frac{\partial}{\partial w^\gamma} \langle x_j^\gamma \rangle \right].
\end{equation}
Bringing everything together, we finally have
\begin{equation}
    S_{ij}^\alpha = - \sum_\gamma \left[ \Gamma_\gamma \langle x_i^\alpha x_j^\gamma\rangle_c + \frac{\partial}{\partial w^\gamma} \langle x_i^\alpha x_j^\gamma \rangle -\langle x_i^\alpha \rangle  \frac{\partial}{\partial w^\gamma} \langle x_j^\gamma \rangle \right] + \langle x_j^\alpha \rangle \frac{\partial}{\partial w^\alpha} \langle x_i^\alpha \rangle.
\end{equation}
When there are no interactions, $\gamma \neq \alpha$, $\langle x_i^\alpha x_j^\gamma \rangle_c = 0$, as well as $\frac{\partial}{\partial w^\gamma} \langle x_i^\alpha \rangle = 0$. As a result, we are simply left with
\begin{equation}
    S_{ij}^\alpha = -\Gamma_\alpha \langle x_i^\alpha x_j^\alpha \rangle_c - \frac{\partial}{\partial w^\alpha} \langle x_i^\alpha x_j^\alpha \rangle_c,
\end{equation}
which is symmetric and identical to the single-agent system, as expected.

\subsection{Fluctuation-Response Relations}
\label{appendix:FDR}

We start from the most general expression for the partition function
\begin{equation}
    Z_N = \int\mathcal{D} \mathbf{x} \, \e^{\beta U (\{\mathbf{x}^\gamma\})} \prod_\gamma \delta (\mathbf{x}^\gamma \cdot \mathbf{p} - w^\gamma),
\end{equation}
and set-out to find a readily measurable expression for $\Gamma_\alpha = \frac{\partial}{\partial w^\alpha} \log Z_N$. Starting from the computation of
\begin{equation}
    \frac{\partial Z_N}{\partial w^\alpha} = \int \mathcal{D} \mathbf{x}\, \e^{\beta U(\{\mathbf{x}^\gamma\})} \frac{\partial}{\partial w^\alpha} \prod_\gamma \delta (\mathbf{x}^\gamma \cdot \mathbf{p} - w^\gamma);
\end{equation}
we employ the generalized chain rule,
\begin{equation}
\begin{aligned}
    \frac{\partial}{\partial w^\alpha} \prod_\gamma \delta (\mathbf{x}^\gamma \cdot \mathbf{p} - w^\gamma) &= \sum_\gamma \frac{\partial}{\partial w^\alpha} \delta (\mathbf{x}^\gamma \cdot \mathbf{p} - w^\gamma) \prod_{\eta \neq \gamma} \delta (\mathbf{x}^\eta \cdot \mathbf{p} - w^\eta) \\
    &= -\frac{1}{p_i} \sum_\gamma \frac{\partial w^\gamma}{\partial w^\alpha} \frac{\partial}{\partial x_i^\gamma} \delta (\mathbf{x}^\gamma \cdot \mathbf{p} - w^\gamma) \prod_{\eta \neq \gamma} \delta (\mathbf{x}^\eta \cdot \mathbf{p} - w^\eta) \quad \forall i.
    \label{equ:chain_rule}
\end{aligned}
\end{equation}
Now, regardless of the assumption on their distribution, the agents' individual budgets are not related, and as such $\frac{\partial w^\gamma}{\partial w^\alpha} = \delta_{\alpha \gamma}$. Reinjecting in the previous expression and integrating by parts, we obtain
\begin{equation}
    \frac{\partial Z_N}{\partial w^\alpha} = \frac{1}{p_i} \int \mathcal{D}\mathbf{x} \, \frac{\partial}{\partial x_i^\alpha} \Big[\e^{\beta U(\{\mathbf{x}^\gamma\})} \Big] \prod_{\eta} \delta (\mathbf{x}^\eta \cdot \mathbf{p} - w),
\end{equation}
which finally gives
\begin{equation}
    \Gamma_\alpha = \frac{\partial}{\partial w^\alpha} \log Z_N = \frac{\beta}{p_i}  \Big \langle  \frac{\partial U}{\partial x_i^\alpha} \Big \rangle \quad \forall i.
\end{equation}

The other terms in the thermodynamic Slutsky matrix expression can be simplified in similar ways. Indeed starting from
\begin{equation}
    \frac{\partial}{\partial w^\gamma} \langle x_i^\alpha x_j^\gamma \rangle = \frac{1}{Z_N} \int \mathcal{D} \mathbf{x} \, x_i^\alpha x_j^\gamma\, \e^{\beta U(\{\mathbf{x}^\eta \})} \frac{\partial}{\partial w^\gamma} \prod_\eta \delta (\mathbf{x}^\eta \cdot \mathbf{p} - w^\eta) - \Gamma_\gamma \langle x_i^\alpha x_j^\gamma\rangle,
\end{equation}
and reusing Eq.~\eqref{equ:chain_rule} and subsequent iterations by parts, it is straightforward to show that
\begin{equation}
    \frac{\partial}{\partial w^\gamma} \langle x_i^\alpha x_j^\gamma \rangle = \frac{1}{p_k} \Big[\delta_{ik} \delta_{\alpha \gamma} \langle x_j^\gamma \rangle + \delta_{jk} \langle x_i^\alpha \rangle + \beta \Big( \Big\langle x_i^\alpha x_j^\gamma \frac{\partial U}{\partial x_k^\gamma} \Big\rangle - \langle x_i^\alpha x_j^\gamma \rangle \Big\langle \frac{\partial U}{\partial x_k^\gamma} \Big\rangle \Big) \Big] \quad \forall k,
\end{equation}
from which we can also derive
\begin{equation} \label{eq:partialwx}
    \frac{\partial}{\partial w^\gamma} \langle x_i^\alpha \rangle = \frac{1}{p_j} \Big[\delta_{ij} \delta_{\alpha \gamma} + \beta \Big( \Big\langle x_i^\alpha \frac{\partial U}{\partial x_j^\gamma} \Big\rangle - \langle x_i^\alpha \rangle \Big\langle \frac{\partial U}{\partial x_j^\gamma} \Big\rangle \Big) \Big] \quad \forall j.
\end{equation}

\subsection{Aggregate Slutsky Matrix}
\label{appendix:Slutsky_aggregate}
We now take
\begin{equation}
    \mathcal{S}_{ij} := \frac{\partial}{\partial p_j} \langle \overline{x}_i \rangle + \langle \overline{x}_j \rangle \frac{\partial}{\partial \overline{w}} \langle \overline{x}_i \rangle,
\end{equation}
with $\overline{w}$ the arithmetic mean over the agents' budgets. Now, the above can simply be rewritten as
\begin{equation}
    \mathcal{S}_{ij} = \frac{1}{N} ÷\sum_\alpha \frac{\partial}{\partial p_j} \langle x_i^\alpha \rangle + \frac{1}{N^2} \sum_{\alpha,\gamma} \langle x_j^\alpha \rangle \frac{\partial}{\partial \overline{w}} \langle x_i^\gamma \rangle.
\end{equation}
The fluctuation-dissipation relation for the first term still holds as before, while for the derivative with respect to the mean budget we have, in general,
\begin{equation}
 \frac{\partial}{\partial \overline{w}}  \langle x_i^\gamma \rangle  =
 \sum_{\eta} \kappa^\eta  \frac{\partial}{\partial {w}^\eta}  \langle x_i^\gamma \rangle,
\end{equation}
with 
\[ 
\kappa^\eta := \frac{\partial {w}^\eta}{\partial \overline{w}}.
\]
As a result, we may write
\begin{equation}
    \frac{1}{N^2} \sum_{\alpha,\gamma} \langle x_j^\alpha \rangle \frac{\partial}{\partial \overline{w}} \langle x_i^\gamma \rangle =  \frac{1}{N^2} \sum_{\alpha,\gamma,\eta} \kappa^\eta \langle x_j^\alpha \rangle \frac{\partial}{\partial w^\eta} \langle x_i^\gamma \rangle = \frac{1}{N} \sum_{\gamma,\eta} \kappa^\eta \overline{x}_j \frac{\partial}{\partial w^\eta} \langle x_i^\gamma \rangle.
\end{equation}
Bringing both expressions together and changing indices appropriately, we find
\begin{equation}
    \mathcal{S}_{ij} =  - \frac1N \sum_{\alpha, \gamma} \left[ \Gamma_\gamma \langle x_i^\alpha x_j^\gamma\rangle_c + \frac{\partial}{\partial w^\gamma} \langle x_i^\alpha x_j^\gamma \rangle_c +  \left(\langle x_j^\gamma \rangle  - \kappa^\gamma {\overline{x}_j} \right) \frac{\partial}{\partial w^\gamma} \langle x_i^\alpha \rangle   \right].
\end{equation}
In the limit $\beta \to \infty$, we have shown that the second term in the sum vanishes. In this case, we therefore recover a symmetric Slutsky matrix whenever $\langle x_j^\gamma \rangle = \kappa^\gamma \overline{x}_j$. This is for example the case when the budgets are proportional to the mean i.e. $w^\gamma = \kappa^\gamma \overline{w}$ and all agents have identical preferences.

Now, in a simple model where $w^\gamma = \kappa^\gamma \overline{w}$, $\forall \gamma$, and all agents have identical preferences, then
\[ 
\langle x_j^\gamma \rangle  \equiv \kappa^\gamma {\overline{x}_j}
\] 
and the second contribution to $\mathcal{S}_{ij}$ identically vanishes, so 
\begin{equation}
    \mathcal{S}_{ij} =  - \frac1N \sum_{\alpha, \gamma} \left[\Gamma_\gamma \langle x_i^\alpha x_j^\gamma\rangle_c + \frac{\partial}{\partial w^\gamma} \langle x_i^\alpha x_j^\gamma \rangle_c\right],
\end{equation}
which is always symmetric when all agents are identical. 

Interesting things may happen when agents are more heterogeneous, or in a model where a change of average wealth does not affect all agents in an affine way. Imagine agents are characterized by a label $\kappa^\alpha \in \mathbb{R}$ such that 
\begin{equation} 
w^\alpha = w_0 F\left(\frac{\kappa^\alpha {w}}{w_0}\right),
\end{equation} 
where $w_0$ is a certain fixed wealth scale, $w$ a varying parameter and $F$ a possibly non-linear function. We must impose  that 
\[ 
\frac{\overline{w}}{w_0} = \frac{1}{N} \sum_\alpha F\left(\frac{\kappa^\alpha {w}}{w_0}\right).
\] 
For example when $F(x)=x$ one finds the above linear model: $w^\alpha=\kappa^\alpha w$ with 
\[ 
w = \frac{\overline{w}}{\overline{\kappa}}.
\]
More generally, one has
\[ 
\frac{\partial\overline{w}}{\partial w} =  \overline{\kappa F'\left(\frac{\kappa {w}}{w_0}\right)}
\] 
and 
\[ 
\frac{\partial{w}^\alpha}{\partial w} = \kappa^\alpha F'\left(\frac{\kappa^\alpha {w}}{w_0}\right),
\] 
so in general 
\[ 
\frac{\partial{w}^\alpha}{\partial \overline{w}} \neq \frac{{w}^\alpha}{\overline{w}}
\] 
and the resulting Slutsky matrix has no longer obvious reasons to be symmetric.

\subsection{Gaussian Fluctuations}
\label{appendix:Gaussian_correls}

We take $\mathbf{x}^{\alpha} - \langle \mathbf{x}^\alpha \rangle = \delta \mathbf{x}^\alpha$ and set out to calculate $\langle x_i^\alpha x_j^\gamma \rangle_c = \langle \delta x_i^\alpha \delta x_j^\alpha \rangle$. At this stage, for simplicity, we combine the $N\times M$ degrees of freedom in the single column vector $\mathbf{v} = [x_1^1,\dots, x_M^1,\dots,x_1^N,\dots,x_M^N]^\top$ such that the correlations of interest are given by
\begin{equation}
    \langle \delta v_{k_1} \delta v_{k_2} \rangle = \frac{\e^{\beta U^*}}{Z_N} \int_{-\infty}^\infty \bigg( \prod_\alpha \frac{\dd \hat{\mu}^\alpha}{2\pi}\bigg) \int_{-\infty}^\infty \bigg(\prod_{k=1}^{M\times N} \dd \delta v_k \bigg) \delta v_{k_1} \delta v_{k_2} \, \e^{\frac{\beta}{2} \delta \mathbf{v}^\top \mathbf{H} \delta \mathbf{v} + i\delta \mathbf{v} \cdot \tilde{\mathbf{p}}},
    \label{equ:covar}
\end{equation}
with the resized price vector $\tilde{\mathbf{p}} \in \mathbb{R}^{M\times N}$, $\tilde{p}_k = \hat{\mu}^\alpha p_i$, $k = M(\alpha - 1) + i$, and where the product of budget constraints has been rewritten using the Fourier representation of the Dirac $\delta$. In this Gaussian approximation, the partition function can be calculated exactly with two consecutive integration by parts,
\begin{equation}
    Z_N = \e^{\beta U^*} \int_{-\infty}^\infty \bigg( \prod_\alpha \frac{\dd \hat{\mu}^\alpha}{2\pi}\bigg) \int_{-\infty}^\infty \bigg(\prod_{k=1}^{M\times N} \dd \delta v_k \bigg) \, \e^{\frac{\beta}{2} \delta \mathbf{v}^\top \mathbf{H} \delta \mathbf{v} + i\delta \mathbf{v} \cdot \tilde{\mathbf{p}}} = \e^{\beta U^*} \sqrt{\frac{(2\pi)^{M\times N + N}}{\det (-\beta \mathbf{H}) \det(-\mathbf{G}/\beta)}},
\end{equation}
with the $N\times N$ matrix $\mathbf{G}$ defined as
\begin{equation}
    G^{\alpha \gamma} = \sum_{i,j} p_i (\mathbf{H}^{-1})_{ij}^{\alpha \gamma} p_j.
\end{equation}
Going back to Eqs.~\eqref{equ:covar}, completing the square in the exponent gives the change of variable $\delta \mathbf{u} = \delta \mathbf{v} + \frac{i}{\beta} \mathbf{H}^{-1} \tilde{\mathbf{p}}$. The integral then reads
\begin{equation}
    \langle \delta v_{k_1} \delta v_{k_2} \rangle = \frac{\e^{\beta U^*}}{Z_N} \int_{-\infty}^\infty \bigg( \prod_\alpha \frac{\dd \hat{\mu}^\alpha}{2\pi}\bigg) \e^{\frac{1}{2\beta}\tilde{\mathbf{p}}^\top \mathbf{H}^{-1} \tilde{\mathbf{p}}} \int_{-\infty}^\infty \bigg(\prod_{k=1}^{M\times N} \dd \delta u_k \bigg) \, \bigg(\delta u_{k_1} \delta u_{k_2} - \frac{1}{\beta^2} (\mathbf{H}^{-1} \tilde{\mathbf{p}})_{k_1} (\mathbf{H}^{-1} \tilde{\mathbf{p}})_{k_2} \bigg) \e^{\frac{\beta}{2} \delta\mathbf{u}^\top \mathbf{H} \delta \mathbf{u}} .
\end{equation}
The first term is straightforward, and simply gives $-\frac{1}{\beta} (\mathbf{H}^{-1})_{k_1 k_2}$ as without the constraint. The second term requires more care, as we have
\begin{equation}
    (\mathbf{H}^{-1} \tilde{\mathbf{p}})_{k_1} = \sum_{\ell_1} (\mathbf{H}^{-1})_{k_1 \ell_1} \tilde{p}_{\ell_1} = \sum_{\ell_1} (\mathbf{H}^{-1})_{k_1 \ell_1} \hat{\mu}^{\alpha(\ell_1)} p_{i(\ell_1)}.
\end{equation}
As a result,
\begin{equation}
    \int_{-\infty}^\infty \bigg( \prod_\alpha \frac{\dd \hat{\mu}^\alpha}{2\pi}\bigg) (\mathbf{H}^{-1} \tilde{\mathbf{p}})_{k_1} (\mathbf{H}^{-1} \tilde{\mathbf{p}})_{k_2} \, \e^{\frac{1}{2\beta} \hat{\bm{\mu}}^\top \mathbf{G} \hat{\bm{\mu}}} = - \beta \sqrt{\frac{(2\pi)^N}{\det(-\mathbf{G}/\beta)}} \sum_{\ell_1,\ell_2} (\mathbf{H}^{-1})_{k_1 \ell_1} p_{i(\ell_1)} (\mathbf{G}^{-1})_{\alpha(\ell_1) \alpha(\ell_2)} (\mathbf{H}^{-1})_{k_2 \ell_2} p_{i(\ell_2)}.
\end{equation}
Bringing everything together, we finally have
\begin{equation}
    \langle \delta v_{k_1} \delta v_{k_2} \rangle = - \frac{1}{\beta} (\mathbf{H}^{-1})_{k_1 k_2} + \frac{1}{\beta} \sum_{\ell_1,\ell_2} (\mathbf{H}^{-1})_{k_1 \ell_1} p_{i(\ell_1)} (\mathbf{G}^{-1})_{\alpha(\ell_1) \alpha(\ell_2)} (\mathbf{H}^{-1})_{k_2 \ell_2} p_{i(\ell_2)},
\end{equation}
which, after replacing with the correct indices, is the expression given in the main text.

\section{Numerical Methods}
\label{appendix:mc}
\subsection{General Idea}

To measure equilibrium properties of the system, we choose to follow a Monte Carlo approach, although a stochastic gradient descent of the global utility (i.e. Langevin dynamics) should be equivalent. The system is first initialized by randomly allocating the $N$ baskets following a uniform distribution. We then employ a Metropolis-Hastings algorithm (see next subsection), as a modification $\Delta \mathbf{x}$ to a randomly selected agent's basket is proposed and accepted with a rate verifying detailed balance \cite{newman1999monte}. To satisfy the non-negativity of the baskets and to correctly sample the small $x_i^\alpha$ region if high concentration occurs, the move is actually  constructed in terms of logarithms. For a randomly selected agent $\gamma$, we take $\log(\mathbf{x}^\gamma + \Delta \mathbf{x}) = \log \mathbf{x}^\gamma + \bm \xi$, with $\xi_i \sim \mathcal{N}(0,1)$, resulting in a log-normally distributed multiplicative noise. The budget constraint is then enforced by a simple rescaling of the resulting basket of goods. Here, it is essential to note that the proposal distribution is not symmetric. As a result, the asymmetry must be accounted for in the acceptance rate, giving in our specific case
\begin{equation*}
    \mathbb{P}(\mathbf{x}^\gamma \to \mathbf{x}^\gamma + \Delta \mathbf{x}) = \min\Big(1,\e^{\beta \Delta U + \sum_i [ \log(x_i^\gamma + \delta x_i) - \log x_i^\gamma]}\Big),
\end{equation*}
with $\Delta U$ the change in the \textit{global} utility\footnote{See Section~\ref{sec:global_vs_individual} for a complete discussion on the consequences of optimizing the global utility.} caused by the move (see next subsection). From each run, average quantities such as $\langle \mathbf{x}^\alpha \rangle$ are measured by taking arithmetic means in algorithmic time once the system is equilibrated.

\subsection{Metropolis-Hastings Acceptance Rate}
\label{appendix:Metropolis}
In its most general form, the Metropolis-Hastings acceptance rate is given by
\begin{equation}
    \mathbb{P}(\mathbf{x} \to \mathbf{y}) = \min\left(1,\frac{q(\mathbf{y}\, | \, \mathbf{x})}{q(\mathbf{x}\, | \,\mathbf{y})} \frac{\pi(\mathbf{y})}{\pi(\mathbf{x})}\right),
\end{equation}
where $\pi(\mathbf{x})$ is the probability density function we wish to sample, and $q(\mathbf{y} \, | \, \mathbf{x})$ is the conditional probability of proposing the state $\mathbf{y}$ given the current state $\mathbf{x}$. In our case, and as usual in statistical mechanics, we simply have $\pi(\mathbf{y})/\pi(\mathbf{x}) = \e^{\beta \Delta U}$. The ratio of conditional probabilities requires more care. Given the noise $y_i = x_i \e^\xi$, with $\xi$ a Gaussian noise, the probability density of which will be noted $\rho$, we have
\begin{equation}
    q(x_i \, | \, y_i) = \int_\mathbb{R} \dd \xi \, \rho(\xi) \, \delta ( y_i - x_i \e^\xi ) = \frac{1}{y_i} \rho\Big( \log \frac{y_i}{x_i} \Big), 
\end{equation}
and very similarly
\begin{equation}
    q(y_i \, | \, x_i) = \frac{1}{x_i} \rho\Big( -\log \frac{y_i}{x_i} \Big).
\end{equation}
Given the symmetry of $\rho$ for zero-mean noise as is the case here, and the fact that all dimensions are statistically independent,
\begin{equation}
    \frac{q(\mathbf{y}\, | \, \mathbf{x})}{q(\mathbf{x}\, | \,\mathbf{y})} = \prod_{i=1}^M \frac{q(y_i \, | \, x_i)}{q(x_i \, | \, y_i)} = \e^{\sum_i \log\big(\frac{y_i}{x_i} \big)},
\end{equation}
yielding the expression in the previous subsection.

\subsection{Computing the Slutsky Matrix}

To compute the entries of  the Slutsky matrix, one can use the fluctuation-response relations derived here, which do not require derivatives. However, out-of-equilibrium effects can become significant, in which case these relations do not hold. For this reason, it is essential to compute as a check the original expression given in Eq.~\eqref{equ:Slutsky_mat_gen}. This is not entirely straightforward, as taking discrete derivatives after time averaging may induce some biases. Instead, we rely on so-called \textit{pathwise derivative estimates}, often used in mathematical finance to compute risk sensitivities \cite{glasserman1991gradient,glasserman2013monte}. In a nutshell, we generate perturbed trajectories based on the original simulation that will use the same random numbers (both for proposed moves and acceptance), and measure finite differences at every step.

\section{Interacting Model}
We consider the global utility
\begin{equation}
    U(\{\mathbf{x}^\alpha\}) = \sum_{\alpha = 1}^N \sum_{i = 1}^M a_i \log x_i^\alpha [1 + c(\overline{x}_i)^k], \quad \overline{x}_i = \frac{1}{N} \sum_{\alpha = 1}^N x_i^\alpha.
\end{equation}

\subsection{Non-interacting Limit -- Canonical Ensemble}
\label{appendix:heterogeneous}
We start by solving the much simplified non-interacting limit, setting $c = 0$. When there are no interactions in general, the canonical partition function entirely decouples and can be written as a product, 
\begin{equation}
    Z_N = \prod_{\alpha = 1}^N z^\alpha, \quad \text{with} \quad z^\alpha = \int_0^\infty \left( \prod_{i=1}^M \dd x_i^\alpha \right)\, \e^{\beta u(\mathbf{x}^\alpha)} \delta(\mathbf{p}\cdot \mathbf{x}^\alpha - w^\alpha).
\end{equation}
Now, plugging in the desired utility function and rewriting the Dirac delta with its integral representation, we have
\begin{equation}
    z^\alpha = \int_0^\infty \left( \prod_{i=1}^M \dd x_i^\alpha \right) \int_{-i\infty}^{i\infty} \frac{\dd \mu}{2\pi i} \, \e^{\beta \sum_i a_i \log x_i^\alpha - \mu \sum_i p_i x_i^\alpha + \mu w^\alpha}.
\end{equation}
This integral now decouples in a product over $i$, and performing the integral on $x_i^\alpha$ with the appropriate change of variable,
\begin{equation}
\begin{aligned}
    z^\alpha &= \int_{-i\infty}^{i\infty} \frac{\dd \mu}{2\pi i} \, \e^{\mu w^\alpha} \prod_i \frac{\Gamma(1+\beta a_i)}{(\mu p_i)^{1+\beta a_i}}\\
    &= \left( \prod_i \frac{\Gamma(1+\beta a_i)}{p_i^{1+\beta a_i}} \right) \mathcal{L}^{-1}_\mu \left\{ \mu^{-\sum_i(1+\beta a_i)} \right\}(w^\alpha)\\
    &= \frac{(w^\alpha)^{\sum_i (1+\beta a_i) - 1}}{\Gamma(M + \beta \sum_i a_i)} \prod_{i=1}^M \frac{\Gamma(1+\beta a_i)}{p_i^{1+\beta a_i}},
\end{aligned}
\end{equation}
where we have used the known inverse Laplace transform of a power law. Taking the product over all agents then gives the result in the text for the partition function.

To find the average $\langle x_i^\alpha \rangle$, we proceed directly, by computing
\begin{equation}
    \langle x_i^\alpha \rangle = \frac{1}{Z_N} \left[\int_0^\infty \left( \prod_{i=1}^M \dd x_i^\alpha \right)\, x_i^\alpha \e^{\beta u(\mathbf{x}^\alpha)} \delta(\mathbf{p}\cdot \mathbf{x}^\alpha - w^\alpha)\right] \prod_{\gamma \neq \alpha} z^{\gamma}
\end{equation}
i.e. exactly as before except for the agent and product considered, for which we must now calculate
\begin{equation}
    \int_0^\infty \dd x_i^\alpha \int_{-i\infty}^{i\infty} \frac{\dd \mu}{2\pi i} \, x_i^\alpha \e^{\beta a_i \log x_i^\alpha - \mu p_i x_i^\alpha + \mu w^\alpha} = \int_{-i\infty}^{i\infty} \frac{\dd \mu}{2\pi i} \, \frac{\Gamma(2+\beta a_i)}{(\mu p_i)^{2+\beta a_i}} = \int_{-i\infty}^{i\infty} \frac{\dd \mu}{2\pi i} \, \frac{1+\beta a_i}{\mu p_i} \frac{\Gamma(1+\beta a_i)}{(\mu p_i)^{1+\beta a_i}},
\end{equation}
where we have used the property $\Gamma(x+1) = x \Gamma(x)$. The integration over $\mu$ is then identical with only the change in the exponent. At this stage, we can plug back everything together, all $\gamma \neq \alpha$, $j\neq i$ terms will cancel out from the $Z_N$ at the denominator, and we are left with
\begin{equation}
    \langle x_i^\alpha \rangle = \frac{w^\alpha}{p_i} \frac{1+\beta a_i}{\sum_k(1 + \beta a_k)}.
\end{equation}
Thanks to this explicit form, the agent-specific Slutsky matrix entries read
\begin{equation}
    S_{ij}^\alpha = \frac{w^\alpha}{p_i p_j} \frac{1+\beta a_i}{\sum_k(1 + \beta a_k)} \left[\frac{1+\beta a_j}{\sum_k(1 + \beta a_k)} - \delta_{ij} \right],
\end{equation}
which, as predicted by thermodynamic relations, is clearly symmetric, and where the heterogeneous budgets only act a a prefactor altering the magnitude, such that we can write $S_{ij}^\alpha = w^\alpha K_{ij}$. The distribution of $S_{ij}^\alpha$ is then simply a rescaling of the original distribution of $w^\alpha$ in the large $N$ limit.



\subsection{Finite Interactions -- Grand-Canonical Ensemble}
\label{appendix:interacting}
To tackle finite interactions, we need to place ourselves in the grand-canonical ensemble, where calculations are possible. Enforcing the value of $\overline{x}_i$ to be the arithmetic mean over all agents with a Dirac distribution expressed in integral form, we have
\begin{equation}
\begin{aligned}
    \mathcal{Z}_N = \int_0^\infty \left( \prod_{\alpha = 1}^N \prod_{i=1}^M \dd x_i^\alpha \right) \int_0^\infty \left( \prod_{i=1}^M \dd \overline{x}_i \right) \int_{-i\infty}^{i\infty} \left( \prod_{i=1}^M \frac{\dd \lambda_i}{2\pi i} \right) \exp\Bigg( &\beta \sum_{\alpha,i} a_i \log x_i^\alpha [1+ c(\overline{x}_i)^k] - \beta \mu \sum_{\alpha,i} x_i^\alpha p_i \\
    &- N\sum_i \overline{x}_i \lambda_i + \sum_{\alpha,i} x_i^\alpha \lambda_i \Bigg).
\end{aligned}
\end{equation}
We may begin by integrating over $x_i^\alpha$. Given the positive support and logarithmic utility, the result is simply a Gamma function for the $N$ identical agents,
\begin{equation}
    \mathcal{Z}_N = \int_0^\infty \left( \prod_{i=1}^M \dd \overline{x}_i \right) \int_{-i\infty}^{i\infty} \left( \prod_{i=1}^M \frac{\dd \lambda_i}{2\pi i} \right) \Bigg( \prod_{i=1}^M \frac{\Gamma(1+\beta a_i[1+c (\overline{x}_i)^k])}{(\beta\mu p_i - \lambda_i)^{1+\beta a_i [1+c(\overline{x}_i)^k]}} \Bigg)^N \, \e^{-N\sum_i \overline{x}_i \lambda_i}.
\end{equation}
The entire integrand can thus be rewritten as an exponential with $N$ as a prefactor,
\begin{equation}
\begin{aligned}
    \mathcal{Z}_N = \int_0^\infty \left( \prod_{i=1}^M \dd \overline{x}_i \right) \int_{-i\infty}^{i\infty} \left( \prod_{i=1}^M \frac{\dd \lambda_i}{2\pi i} \right) \, \exp \Big( -N \sum_i \Big[& \overline{x}_i \lambda_i + (1+\beta a_i [1+c(\overline{x}_i)^k ]) \log(\beta \mu p_i - \lambda_i)\\
    &- \log \Gamma(1+\beta a_i [1+c(\overline{x}_i)^k] \Big] \Big),
\end{aligned}
\end{equation}
a form which naturally points to the saddle-point approximation as $N\to \infty$. Indeed, the integral over $\lambda_i$ above may first be be rewritten as approximated as
\begin{equation}
    \int_{-i\infty}^{i\infty} \left( \prod_{i=1}^M \frac{\dd \lambda_i}{2\pi i} \right) \, \e^{-N g(\overline{\mathbf{x}}, \bm{\lambda})} \sim \sqrt{\frac{2\pi}{N |\det \mathbf{H}^*|}} \, \e^{-N g(\overline{\mathbf{x}},\bm{\lambda}^*)},
\end{equation}
where $\bm{\lambda}^*$ is a minimum of $g$, and $\mathbf{H}^*$ is the Hessian with respect to $\lambda$ evaluated at that point. At the saddle, we have
\begin{equation}
  \frac{\partial g}{\partial \lambda_i} = \overline{x}_i - \frac{1+\beta a_i [1+c(\overline{x}_i)^k]}{\beta \mu p_i - \lambda_i^*} = 0 \quad \Rightarrow \quad \lambda_i^* = \beta \mu p_i - \frac{1}{\overline{x}_i}(1+\beta a_i [1+c(\overline{x}_i)^k]),
    \label{equ:saddle_1}
\end{equation}
and the Hessian is entirely diagonal and given by
\begin{equation}
    H_{ij} = \frac{1+\beta a_i[1+c(\overline{x}_i)^k]}{(\beta \mu p_i - \lambda_i)^2} \delta_{ij} \quad \Rightarrow \quad |\det \mathbf{H}^*| = \prod_{i=1}^M \frac{(\overline{x}_i)^2}{1+\beta a_i[1+c(\overline{x}_i)^k]}.
\end{equation}
As a result, the partition function can finally be expressed in the desired form,
\begin{equation}
    \mathcal{Z}_N = \int_0^\infty \dd \overline{\mathbf{x}} \, \e^{-\beta F(\overline{\mathbf{x}})},
\end{equation}
with $F(\overline{\mathbf{x}}) = N f(\overline{\mathbf{x}})$, and where from the above equations we have
\begin{equation}
\begin{aligned}
    \beta f(\overline{\mathbf{x}}) =& \sum_i  \big[ \beta \mu p_i \overline{x}_i - (1 + \beta a_i[1 + c(\overline{x}_i)^k])[1 + \log \overline{x}_i - \log(1+\beta a_i [1+c(\overline{x}_i)^k])] - \log \Gamma(1+\beta a_i [1+c(\overline{x}_i)^k])\big] \\
     & \qquad \qquad \underbrace{+ \frac{1}{2} \frac{\log N}{N} + \frac{1}{N}\left[ \sum_i \Big[\log \overline{x}_i - \frac{1}{2} \log(1+\beta a_i[1+c(\overline{x}_i)^k]) \Big] - \frac{1}{2} \log 2\pi \right]}_{o(1)},
\end{aligned}
\label{equ:f_GC}
\end{equation}
which is completely decoupled in between degrees of freedom. Keeping only $\mathcal{O}(1)$ terms, we find the expression given in the main text that we will use onward.

In the thermodynamic limit, the state of the system can once again be determined through the saddle-point approximation, as the system will reach a minimum of $f$ with overwhelming likelihood. This amounts to solving
\begin{equation}
    \frac{\partial f}{\partial \overline{x}_i} = -\frac{1}{\overline{x}_i^*} \left( \frac{1}{\beta} - \mu p_i \overline{x}_i^* + a_i\left[ 1 + c (\overline{x}_i^*)^k (1+k[\psi(1+\beta a_i[1+c(\overline{x}_i^*)^k]) - \log (1+\beta a_i[1+c(\overline{x}_i^*)^k]) + \log \overline{x}_i^*]) \right] \right) = 0,
\end{equation}
with $\psi$ the digamma function. To determine the nature of stationary points, we will also require
\begin{equation}
    \begin{aligned}
        \frac{\partial^2 f}{\partial \overline{x}_i^2} = \frac{1}{(\overline{x}_i)^2} & \Bigg[\frac{(1/\beta + a_i[1-c(k-1)(\overline{x}_i)^k])^2}{1/\beta + a_i[1+c(\overline{x}_i)^k]} - k c a_i (\overline{x}_i)^k \Big( \beta k c a_i (\overline{x}_i)^k \psi^{(1)}(1+\beta a_i[1+c(\overline{x}_i)^k]) \\
        & +[k-1][\psi(1+\beta a_i[1+c(\overline{x}_i)^k]) - \log(1+\beta a_i[1+c(\overline{x}_i)^k]) + \log \overline{x}_i] \Big)  \Bigg],
    \end{aligned}
    \label{equ:f''_GC}
\end{equation}
where $\psi^{(1)}$ is the first polygamma function. Remarkably, the second derivative is not dependent on the chemical potential $\mu$. Importantly, one can evaluate the partition function by steepest descent to show that, consistent with intuition,
\begin{equation}
    \langle x_i \rangle = -\frac{1}{N \beta \mu} \frac{\partial}{\partial p_i}\log \mathcal{Z} = \overline{x}_i^*,
    \label{equ:x_i_GC_saddle}
\end{equation}
when $N\to \infty$, straight from Eq.~\eqref{equ:f_GC} as all derivatives other that those with respect to $p_i$ are zero at the saddle by construction. 

\paragraph{$c = 0$ solution} Taking $c = 0$, we easily recover the unconcentrated solution 
\begin{equation}
    \overline{x}_i^* = \frac{w}{p_i} \frac{1+\beta a_i}{\sum_k (1+ \beta a_k)}
\end{equation}

\paragraph{$\beta \to \infty$ limit} In order to better understand the influence of the different terms on the equilibrium solution of the system, we start by studying the zero temperature, or fully rational, limit of the system. Knowing the asymptotics of the digamma function $\psi(z) \sim \log z - \frac{1}{2z}$, we obtain
\begin{equation}
    a_i[1 + c(\overline{x}_i^*)^k (1+ k\log \overline{x}_i^*)] = \mu p_i \overline{x}_i^*, \qquad \sum_i \overline{x}_i^* p_i = w,
\end{equation}
i.e. a set of $M + 1$ equations for the $M$ variables $\overline{x}_i^*$ and the correct value of $\mu$ for the given budget $w$. To obtain the solution as a function of $c$, one can numerically solve the system of equations. 

Knowing what the uniform solution is, we may then identify where the phase transition occurs, i.e. what the value of $c_\infty$ is, by studying the stability of the free energy at that point, as it should go from being a minimum to a maximum at the transition. Now in this limit, we have
\begin{equation}
    \frac{\partial^2 f}{\partial \overline{x}_i ^2} = \frac{a_i}{(\overline{x}_i)^2} \left[ \frac{(1-c(k-1)(\overline{x}_i)^k)^2}{1+c(\overline{x}_i)^k} - kc(\overline{x}_i)^k\left( \frac{kc(\overline{x}_i)^k}{1+c(\overline{x}_i)^k} + (k-1) \log \overline{x}_i \right) \right].
\end{equation}
Setting to zero at the saddle, and solving for $c_\infty$ finally yields
\begin{equation}
    \frac{1}{c_\infty} = \underset{\overline{x}_i^*}{\max} \; (\overline{x}_i^*)^k\left[ 2k - 1 + k(k-1)\log \overline{x}_i^* \right].
\end{equation}
In the case where all products are equivalent, we recover the expression in the main text.

\paragraph{Finite $\beta$}
In the general case, the equations for $\overline{x}_i^*$ are much harder to manipulate due to the presence of the highly nonlinear logarithm and digamma functions. However, the method outlined to obtain the critical value of $c$ remains valid: the $c = 0$ solution is to be plugged into the second derivative that is set to zero. Solving numerically (Eq.~\eqref{equ:f''_GC}) then yields the critical line in $(c,\beta)$ shown in the main text and in Fig. \ref{fig:Transition_betafinite}.


\subsection{Slutsky Matrix in the $\beta \to \infty$ Limit}
\label{appendix:SlutskyZeroT}
We wish to employ the previously devised Gaussian approximation for the Slutsky matrix at $\beta \to \infty$ for our interacting model. To do so means evaluating the ``thermodynamic'' expression of the Slutsky matrix that requires computing the cross-agent term $\langle x_j^\gamma \rangle \frac{\partial}{\partial w^\gamma} \langle x_i^\alpha \rangle$. For our specific choice of utility,
\begin{equation}
    \frac{\partial U}{\partial x_j^\gamma} = \frac{a_j}{x_j^\gamma} [1+c(\overline{x}_j)^k] + k c a_j  (\overline{x}_j)^{k-1} \overline{\log x_j},
\end{equation}
with overlines indicating arithmetic averages over the $N$ agents. As we are interested in the $N \to \infty$ behavior, we can first notice that in the vicinity of the maximum
\begin{equation}
    \overline{x}_i = \overline{x}_i^* + \frac{1}{N} \sum_\alpha \delta x_i^\alpha = \overline{x}_i^* + \mathcal{O}\bigg( \frac{1}{\sqrt{N}} \bigg)
\end{equation}
by virtue of the central limit theorem. Similarly,
\begin{equation}
    \overline{\log x_i} = \frac{1}{N} \sum_\alpha \log(\overline{x}_i^* + \delta x_i^\alpha) = \log \overline{x}_i^* + \frac{1}{N} \sum_\alpha \bigg[ \frac{\delta x_i^\alpha}{\overline{x}_i^*} - \frac{1}{2} \bigg(\frac{\delta x_i^\alpha}{\overline{x}_i^*} \bigg)^2 + o\bigg(\frac{1}{\beta}\bigg) \bigg] = \log \overline{x}_i^* + \mathcal{O}\bigg( \frac{1}{\sqrt{N}} \bigg) + \mathcal{O}\bigg( \frac{1}{\beta} \bigg).
\end{equation}
To leading order, we therefore have
\begin{equation}
    \frac{\partial U}{\partial x_j^\gamma} = \frac{a_j}{\overline{x}_j^* + \delta x_j^\gamma} [1 + c (\overline{x}_j^*)^k] + k c a_j (\overline{x}_j^*)^{k-1} \log \overline{x}_j^* + \mathcal{O}\bigg( \frac{1}{\sqrt{N}} \bigg).
\end{equation}
As a result, developing to the second order in $\delta x_j^\gamma$
\begin{equation}
    \Big\langle x_i^\alpha \frac{\partial U}{\partial x_j^\gamma} \Big\rangle - \langle x_i^\alpha \rangle \Big\langle \frac{\partial U}{\partial x_j^\gamma} \Big\rangle = -\frac{a_j}{(\overline{x}_j^*)^2} \langle  x_i^\alpha  x_j^\gamma \rangle_c [1+c(\overline{x}_j^*)^k] + \mathcal{O}\bigg( \frac{1}{\sqrt{N}} \bigg),
\end{equation}
so plugging back this expression in the fluctuation-response relation gives 
\begin{equation}
    \frac{\partial}{\partial w^\gamma} \langle x_i^\alpha \rangle = \frac{1}{p_j} \Big(\delta_{ij} \delta_{\alpha \gamma} - \beta \frac{a_j}{(\overline{x}_j^*)^2} \langle  x_i^\alpha  x_j^\gamma \rangle_c  [1+c(\overline{x}_j^*)^k] \Big) \quad \forall j.
\end{equation}
Given the Gaussian fluctuations scale as $\langle x_i^\alpha  x_j^\gamma \rangle_c \sim {\beta}^{-1}$, this term will clearly have a non-negligible contribution to the Slutsky matrix at $\beta \to \infty$.

Using the fluctuation-response relation to compute the first term in the Slutsky matrix,
\begin{equation}
\begin{aligned}
    \Gamma_\gamma &= \frac{\beta}{p_\ell} \bigg \langle \frac{\partial U}{\partial x_\ell^\gamma} \bigg\rangle \quad \forall \ell \\
    &= \beta \frac{a_\ell}{p_\ell \overline{x}_\ell^*}\bigg[ 1 + c (\overline{x}_\ell^*)^k(1+k\log \overline{x}_\ell^*) + \mathcal{O}\bigg( \frac{1}{\beta} \bigg) \bigg],
\end{aligned}
\end{equation}
we can finally write, choosing $\ell=j$
\begin{equation}
    S_{ij}^\alpha = - \beta \sum_\gamma  \langle  x_i^\alpha  x_j^\gamma \rangle_c \frac{a_j}{p_j \overline{x}_j^*} \left( k c (\overline{x}_j^*)^k \log \overline{x}_j^* + \delta_{\alpha \gamma} [1+c(\overline{x}_j^*)^k] \right).
\end{equation}

There now remains to compute the pairwise correlations, which requires us to invert the Hessian of the utility function. In our specific case 
\begin{equation}
    H_{ij}^{\alpha \gamma} = \frac{\partial ^2 U}{\partial x_i^\alpha \partial x_j^\gamma} = -\bigg[ \frac{a_i}{(x_i^\gamma)^2}[1+c(\overline{x}_i)^k] \delta_{\alpha \gamma} - \frac{kc}{N}a_i (\overline{x}_i)^{k-1} \bigg( \frac{1}{x_i^\alpha} + \frac{1}{x_i^\gamma}\bigg) + \frac{k(k-1)c}{N} a_i \overline{\log x_i} (\overline{x}_i)^{k-2} \bigg] \delta_{ij}.
\label{equ:Hessian_full}
\end{equation}
Due to the homogeneity in between agents, this $M\times N$ matrix has the very simple structure
\begin{equation}
    \mathbf {H} = {\begin{bmatrix}\mathbf {A} &0&\cdots &0\\0&\mathbf {A}&\cdots &0\\\vdots &\vdots &\ddots &\vdots \\0&0&\cdots &\mathbf {A}\end{bmatrix}}
    +
    {\begin{bmatrix}\mathbf {B}&\mathbf {B}&\cdots &\mathbf {B}\\\mathbf {B}&\mathbf {B}&\cdots &\mathbf {B}\\\vdots &\vdots &\ddots &\vdots \\\mathbf {B}&\mathbf {B}&\cdots &\mathbf {B}\end{bmatrix}},
\end{equation}
with $\mathbf{A},\mathbf{B} \in \mathbb{R}^{M\times M}$, both having the additional simplification of being symmetric matrices. We take the \textit{ansatz} that the inverse has an identical structure,
\begin{equation}
    \mathbf {H}^{-1} = {\begin{bmatrix}\mathbf {F} &0&\cdots &0\\0&\mathbf {F}&\cdots &0\\\vdots &\vdots &\ddots &\vdots \\0&0&\cdots &\mathbf {F}\end{bmatrix}}
    +
    {\begin{bmatrix}\mathbf {D}&\mathbf {D}&\cdots &\mathbf {D}\\\mathbf {D}&\mathbf {D}&\cdots &\mathbf {D}\\\vdots &\vdots &\ddots &\vdots \\\mathbf {D}&\mathbf {D}&\cdots &\mathbf {D}\end{bmatrix}}.
\end{equation}
As such, we have
\begin{equation}
    \mathbf{H} \mathbf{H}^{-1} = {\begin{bmatrix}\mathbf{A}\mathbf{F} &0&\cdots &0\\0&\mathbf{A}\mathbf{F}&\cdots &0\\\vdots &\vdots &\ddots &\vdots \\0&0&\cdots &\mathbf{A}\mathbf{F}\end{bmatrix}}
    +
    {\begin{bmatrix}\mathbf {E}&\mathbf {E}&\cdots &\mathbf {E}\\\mathbf {E}&\mathbf {E}&\cdots &\mathbf {E}\\\vdots &\vdots &\ddots &\vdots \\\mathbf {E}&\mathbf {E}&\cdots &\mathbf {E}\end{bmatrix}}, \qquad \mathbf{E} = \mathbf{A} \mathbf{D} + \mathbf{B} \mathbf{F} + N \mathbf{B} \mathbf{D},
\end{equation}
obtained from the multiplication of these simple block matrices. As a result, setting $\mathbf{H} \mathbf{H}^{-1} = \mathbf{I}$ amounts to
\begin{equation}
    \mathbf{F} = \mathbf{A}^{-1}, \qquad \mathbf{D} = -(\mathbf{A} + N \mathbf{B})^{-1} \mathbf{B} \mathbf{A}^{-1}.
\end{equation}
Going back to Eq.~\eqref{equ:Hessian_full} (evaluated at the maximum $\overline{\mathbf{x}}^*$), we have
\begin{equation}
    A_{ij} = - a_i \bigg( \frac{1+c(\overline{x}_i^*)^k}{(\overline{x}_i^*)^2} + \frac{k c}{N} (\overline{x}_i^*)^{k-2}[2 + (k-1) \log \overline{x}_i^*]\bigg) \delta_{ij}
\end{equation}
and 
\begin{equation}
    B_{ij} = \frac{k c}{N} a_i (\overline{x}_i^*)^{k-2}[2 + (k-1) \log \overline{x}_i^*]\delta_{ij}.
\end{equation}
The diagonal matrix can be inverted (almost) exactly in the large $N$ limit, yielding
\begin{equation}
    F_{ij} = - \frac{\delta_{ij}}{a_i} \frac{(\overline{x}_i^*)^2}{1+c(\overline{x}_i^*)^k} \bigg[ 1 - \frac{kc}{N} \frac{(\overline{x}_i^*)^k}{1+c(\overline{x}_i^*)^k}(2 + (k-1)\log \overline{x}_i^*) + \mathcal{O}\bigg(\frac{1}{N^2}\bigg) \bigg],
\end{equation}
and similarly at the leading order
\begin{equation}
    [(\mathbf{A} + N \mathbf{B})^{-1}]_{ij} = - \frac{\delta_{ij}}{a_i} \frac{(\overline{x}_i^*)^2}{1+c(\overline{x}_i^*)^k[1-k(2+(k-1)\log \overline{x}_i^*)]} \bigg[ 1 - \frac{k c}{N} \frac{(\overline{x}_i^*)^k (2 + (k-1)\log \overline{x}_i^*)}{1+c(\overline{x}_i^*)^k[1-k(2+(k-1)\log \overline{x}_i^*)]} \bigg],
\end{equation}
finally giving
\begin{equation}
    D_{ij} = - \frac{k c}{N} \frac{\delta_{ij}}{a_i} \frac{(\overline{x}_i^*)^{k+2}(2+(k-1)\log \overline{x}_i^*)}{(1+c(\overline{x}_i^*)^k)(1+c(\overline{x}_i^*)^k[1-k(2+(k-1)\log \overline{x}_i^*)])} + \mathcal{O}\bigg(\frac{1}{N^2} \bigg).
\end{equation}

Using the inverse of the Hessian, we can then compute the correlations of interest. We first need
\begin{equation}
    G^{\alpha \gamma} = \sum_{i,j} p_i (\mathbf{H}^{-1})_{ij}^{\alpha \gamma} p_j,
\end{equation}
which can clearly be split in diagonal and off-diagonal parts, therefore the matrix has a similar structure to the Hessian, but reduced in dimension. We have $\mathbf{G} = a \mathbf{I} + b \mathbf{E}$ with $\mathbf{I}$ and $\mathbf{E}$ respectively and identity matrix and a matrix full of ones, both of dimension $N \times N$. Taking the same ansatz as before but with scalars, we have
\begin{equation}
    \mathbf{G}^{-1} = \frac{1}{a} \mathbf{I} + d \mathbf{E}, \qquad \mathrm{with} \quad d =  -\frac{b}{a(a + Nb)}, \quad a = \mathbf{p}^\top \mathbf{F} \mathbf{p}, \quad b = \mathbf{p}^\top \mathbf{D} \mathbf{p}
\end{equation}

Bringing everything together, we finally find,
\begin{equation}
\begin{aligned}
    \langle x_i^\alpha x_j^\gamma \rangle_c &= -\frac{1}{\beta}(\delta_{\alpha \gamma} F_{ii} + D_{ii}) \delta_{ij} + \frac{p_i p_j}{\beta} [F_{ii} F_{jj}(\delta_{\alpha \gamma}a^{-1} + d) + (a^{-1} + Nd)(F_{ii} D_{jj} + D_{ii} F_{jj} + N D_{ii} D_{jj})]\\
    &= \frac{1}{\beta}\bigg[\varphi_{ij} \delta_{\alpha \gamma} + \frac{1}{N}\psi_{ij} \bigg]
\end{aligned}
\end{equation}
with
\begin{equation}
    \varphi_{ij} = F_{ii} \delta_{ij} - a^{-1} p_i p_j F_{ii} F_{jj} \quad \mathrm{and} \quad \frac{1}{N} \psi_{ij} = D_{ii} \delta_{ij} - p_i p_j [d F_{ii} F_{jj} + (a^{-1} + Nd)(F_{ii} D_{jj} + D_{ii} F_{jj} + N D_{ii} D_{jj})]
\end{equation}
the non-interacting and interacting parts respectively. Given that $D_{ii} \sim \frac{1}{N}$ and $d \sim \frac{1}{N}$, it is clear that we have $\varphi_{ij} = \mathcal{O}(1)$ and $\psi_{ij} = \mathcal{O}(1)$, as expected.

Ultimately, in the $N \to \infty$, $\beta \to \infty$ limits, the Slutsky matrix is thus given by
\begin{equation}
    S_{ij}^\alpha = S_{ij} = - \frac{a_j}{p_j} \left[ kc(\varphi_{ij} + \psi_{ij}) (\overline{x}_j^*)^{k-1} \log \overline{x}_j^* + [1+c(\overline{x}_j^*)^k] \varphi_{ij} \right].
\end{equation}
The evolution of this analytical expression for some $\mathbf{a},\mathbf{p}$ with $c$, compared with numerical simulations, is shown in Figure \ref{fig:S_full} of the main text.


\subsection{Equivalence of Ensembles}
\label{appendix:ensembles}
We wish to compute
\begin{equation}
    \sigma^2 = \Big \langle \Big( \sum_i x_i^\alpha p_i - w \Big)^2 \Big \rangle = \Big \langle \Big( \sum_i x_i^\alpha p_i \Big)^2 \Big \rangle - w^2.
\end{equation}
To this end, we take an agent-specific perturbation to the chemical potential $\mu \to \mu + \delta \mu^\alpha$, playing the role of an external field in conventional statistical physics. The grand-canonical partition function then reads
\begin{equation}
    \mathcal{Z}_N = \int_0^\infty \bigg( \prod_{\alpha,i} \dd x_i^\alpha \bigg)\, \e^{\beta[U(\{\mathbf{x}\}) - \mu \sum_{\alpha,i} x_i^\alpha p_i - \sum_{\alpha,i} \delta \mu^\alpha x_i^\alpha p_i ]},
\end{equation}
such that
\begin{equation}
    \frac{1}{\mathcal{Z}_N} \frac{\partial \mathcal{Z}_N}{\partial \delta \mu^\alpha} = -\beta \Big \langle \sum_i x_i^\alpha p_i \Big \rangle = -\beta w \quad \mathrm{and} \quad \frac{1}{\mathcal{Z}_N} \frac{\partial^2 \mathcal{Z}_N}{\partial {\delta \mu^\alpha}^2} = \beta^2 \Big \langle \Big( \sum_i x_i^\alpha p_i \Big)^2 \Big \rangle,
\end{equation}
giving
\begin{equation}
    \sigma^2 = \frac{1}{\beta^2} \frac{\partial^2}{\partial {\delta \mu^\alpha}^2} \log \mathcal{Z}_N.
\end{equation}
As a result, assuming $\delta \mu^\alpha$ to be small, we need to calculate the free energy to the second order in this perturbation. Following the previous calculation, we have
\begin{equation}
    \mathcal{Z}_N = \int_0^\infty \left( \prod_{i=1}^M \dd \overline{x}_i \right) \int_{-i\infty}^{i\infty} \left( \prod_{i=1}^M \frac{\dd \lambda_i}{2\pi i} \right)  \prod_{\alpha,i} \frac{\Gamma(1+\beta a_i[1+c (\overline{x}_i)^k])}{(\beta\mu p_i + \beta \delta\mu^\alpha p_i - \lambda_i)^{1+\beta a_i [1+c(\overline{x}_i)^k]}} \, \e^{-N\sum_i \overline{x}_i \lambda_i}.
\end{equation}
Expanding at the second order in the perturbation, we may write
\begin{equation}
    \mathcal{Z}_N = \int_0^\infty \left( \prod_{i=1}^M \dd \overline{x}_i \right) \int_{-i\infty}^{i\infty} \left( \prod_{i=1}^M \frac{\dd \lambda_i}{2\pi i} \right) \, \e^{-N[g_0(\overline{\mathbf{x}},\bm{\lambda}) + g_1(\overline{\mathbf{x}},\bm{\lambda}) \overline{\delta \mu} + g_2(\overline{\mathbf{x}},\bm{\lambda}) \overline{\delta\mu^2}]},
\end{equation}
to be evaluated through a saddle-point approximation, with $g_0(\overline{\mathbf{x}},\bm{\lambda})$ corresponding to the exponent in the unperturbed expression previously studied,
\begin{equation}
    g_1(\overline{\mathbf{x}},\bm{\lambda}) = \sum_i \frac{p_i}{\mu p_i - \lambda_i/\beta} (1+\beta a_i[1+c(\overline{x}_i)^k])
\end{equation}
and 
\begin{equation}
    g_2(\overline{\mathbf{x}},\bm{\lambda}) = - \frac{1}{2} \sum_i \frac{p_i^2}{(\mu p_i - \lambda_i/\beta)^2} (1+\beta a_i[1+c(\overline{x}_i)^k]),
\end{equation}
while the previously introduced overline notation implies an arithmetic average over agents, i.e. $\overline{\delta \mu} = \frac{1}{N} \sum_\alpha \delta \mu^\alpha$. Now before actually using these expressions, one may use the properties at the saddle to identify the necessary terms to evaluate at the leading order. Indeed, at the saddle in $\lambda$, which decouples in $i$ so can be treated in one dimension for now, we will have
\begin{equation}
    \bm{\lambda} = \bm{\lambda}^* + \bm{\theta}_1 \overline{\delta \mu} + \bm{\theta}_2 \overline{\delta \mu^2} + \mathcal{O}(\delta\mu^3)
\end{equation}
solution to
\begin{equation}
    \frac{\partial g_0}{\partial \lambda_i} + \frac{\partial g_1}{\partial \lambda_i} \overline{\delta \mu} + \frac{\partial g_2}{\partial \lambda_i} \overline{\delta \mu^2} = 0,
\end{equation}
and thus where $\lambda^*$ corresponds to the previously calculated unperturbed solution. Now to evaluate the integrand at the saddle, we may expand all expressions to the second order, yielding
\begin{equation}
    \mathcal{Z}_N = \int_0^\infty \bigg( \prod_i \dd \overline{x}_i \bigg) \, \e^{-N \beta f(\overline{\mathbf{x}})},
\end{equation}
with 
\begin{equation}
    \beta f(\overline{\mathbf{x}}) = g_0(\overline{\mathbf{x}},\bm{\lambda}^*) + \frac{1}{2} \bm{\theta}_1^\top \mathbf{H}_\lambda(g_0(\overline{\mathbf{x}},\bm{\lambda}^*)) \bm{\theta}_1 (\overline{\delta \mu})^2 + g_1(\overline{\mathbf{x}},\bm{\lambda}^*) \overline{\delta \mu} + g_2(\overline{\mathbf{x}},\bm{\lambda}^*) \overline{\delta \mu^2},
\end{equation}
with $\mathbf{H}_\lambda(g_0(\overline{\mathbf{x}},\bm{\lambda}^*))$ the Hessian of $g_0$ with respect to the $\lambda_i$ evaluated at the saddle at the leading order, which is diagonal in our decoupled problem, i.e.
\begin{equation}
    \mathbf{H}_\lambda(g_0(\overline{\mathbf{x}},\bm{\lambda}^*)) = \operatorname{diag} \frac{\partial^2 g_0}{\partial \lambda_i^2} \bigg \rvert_{\mathbf{\overline{x},\bm{\lambda}^*}}.
\end{equation}
We may now evaluate the remaining integrals over $\overline{x}_i$ by taking a new saddle, found at
\begin{equation}
    \overline{\mathbf{x}} = \overline{\mathbf{x}}^* + \bm{\kappa}_1 \overline{\delta \mu} + \bm{\kappa}_2 \overline{\delta \mu^2} + \bm{\kappa}_3 (\overline{\delta \mu})^2,
\end{equation}
the solution to the set of equations $\frac{\partial f}{\partial \overline{x}_i} = 0$. Once again expanding all solutions about the saddle, we finally find
\begin{equation}
    \beta f(\overline{x}) = g_0(\overline{\mathbf{x}}^*,\bm{\lambda}^*) + \frac{1}{2}\Big[ \bm{\theta}_1^\top \mathbf{H}_\lambda(g_0(\overline{\mathbf{x}}^*,\bm{\lambda}^*)) \bm{\theta}_1 + \bm{\kappa}_1^\top \mathbf{H}_{\overline{x}}(g_0(\overline{\mathbf{x}}^*,\bm{\lambda}^*)) \bm{\kappa}_1 \Big] (\overline{\delta \mu})^2 + g_1(\overline{\mathbf{x}}^*,\bm{\lambda}^*) \overline{\delta \mu} + g_2(\overline{\mathbf{x}}^*,\bm{\lambda}^*) \overline{\delta \mu^2}.
\end{equation}
However, at this stage, one may notice that
\begin{equation}
    \frac{\partial^2}{\partial {\delta \mu^\alpha}^2} (\overline{\delta \mu})^2 = \frac{2}{N^2},
\end{equation}
whereas
\begin{equation}
    \frac{\partial^2}{\partial {\delta \mu^\alpha}^2} \overline{\delta \mu^2} = \frac{2}{N}.
\end{equation}
At the leading order in $N$, we therefore simply have to evaluate $g_2$ at the original saddle. The result then reads
\begin{equation}
    \sigma^2 = \frac{2}{\beta^2} g_2(\overline{\mathbf{x}}^*,\bm{\lambda}^*) = \frac{1}{\beta^2} \sum_i \frac{p_i^2}{(\mu p_i - \lambda_i^*/\beta)^2} (1+\beta a_i[1+c(\overline{x}_i^*)^k]) + \mathcal{O}\bigg(\frac{1}{N} \bigg).
\end{equation}
Plugging in the known expression of $\lambda_i^*$ at the original saddle, we finally obtain at the thermodynamic limit
\begin{equation}
    \sigma^2 = \sum_i \frac{(\overline{x}_i^* p_i)^2}{1+\beta a_i[1+c(\overline{x}_i^*)^k]}.
\end{equation}
As expected, this quantity vanishes for $\beta \to \infty$, as well as for $c\to \infty$ $\forall \beta > 0$. In the $c < c_\mathrm{crit}$ region, plugging in the non-condensed solution $\overline{x}_i^* = \frac{w}{p_i} \frac{1 + \beta a_i}{\sum_i(1+\beta a_i)}$ yields
\begin{equation}
    \sigma^2 = \frac{w^2}{\sum_i (1+\beta a_i)} = \mathcal{O}\bigg( \frac{1}{M} \bigg).
\end{equation}

\section{A Hamiltonian Utility Function}
\label{appendix:Hamiltonian_U}

Suppose we now take
\begin{equation}
    U(\{\mathbf{x}^\alpha\}) = \sum_{i,\alpha} a_i \log x_i^\alpha + \frac12 \sum_{\substack{i,\alpha,\gamma \\ \gamma \neq \alpha}}  J_i^{\alpha \gamma} (x_i^\alpha)^\rho (x_i^\gamma)^\rho,
\end{equation}
with {\it symmetric} interactions $J_i^{\gamma \alpha} = J_i^{\alpha \gamma}$, $\forall i, \alpha, \gamma$ and $0 < \rho < 1$. In this case, we then have
\begin{equation}
    \frac{\partial U}{\partial x_i^\alpha} = \frac{a_i}{x_i^\alpha} \Big[ 1 + \rho (x_i^\alpha)^{\rho-1} \sum_{\gamma \neq \alpha} J_i^{\alpha \gamma} (x_i^\gamma)^\rho \Big],
\end{equation}
which is equal to the ``selfish'' derivative of utility of agent $\alpha$ defined as
\begin{equation}
    u_\alpha(\{\mathbf{x}^\alpha\}) = \sum_{i} a_i \log x_i^\alpha +  \sum_{\substack{i,\gamma \\ \gamma \neq \alpha}}  J_i^{\alpha \gamma} (x_i^\alpha)^\rho (x_i^\gamma)^\rho.
\end{equation} 
(Note the factor $\frac12$ difference between the definition of $U$ and that of $u_\alpha$). 
This means that in a such a model, the decision-making process based on a purely individualistic change of utility leads to an equilibrium distribution given by the Boltzmann-Gibbs weight $\exp(\beta U)$ (see \cite{grauwin2009competition} for a similar discussion).

We expect such a Hamiltonian model to be qualitatively similar to the one studied in this work. To check that this is the case, we may start by taking the \textit{mean-field} case $J_i^{\alpha \gamma} = a_i J/N$ and the $\beta \to \infty$ limit. Introducing the Lagrange multipliers $\mu^\alpha$, the Lagrangian to minimize is
\begin{equation}
    \mathcal{L}(\{ \mathbf{x}^\alpha \},\{ \bm{\mu}^\alpha \}) = \sum_{i,\alpha} a_i \log x_i^\alpha + J \sum_{i,\alpha} a_i (x_i^\alpha)^\rho \overline{x_i^\rho} - \sum_{\alpha} \mu^\alpha \Big( \sum_i p_i x_i^\alpha - w^\alpha \Big),
\end{equation}
where the overline again means an average over all agents. This yields the following $N\times(M+1)$ equations
\begin{eqnarray}
\frac{\partial \mathcal{L}}{\partial x_i^\alpha} &=& \frac{a_i}{x_i^\alpha} \Big[ 1 + J\rho (x_i^\alpha)^\rho \overline{x_i^\rho} \Big] - \mu^\alpha p_i = 0\\
\frac{\partial \mathcal{L}}{\partial \mu^\alpha} &=& \sum_i p_i x_i^\alpha - w^\alpha = 0.
\end{eqnarray}
Now assuming that all agents have identical budgets, the maximum will be given by $x_i^\alpha = x_i^* \, \forall \alpha$. The solution then satisfies
\begin{equation}
    x_i^* = \frac{w}{p_i} \frac{a_i[1+J\rho (x_i^*)^{2\rho}]}{\sum_j a_j[1+J\rho (x_j^*)^{2\rho}]}.
\end{equation}
In the non-interacting case $J = 0$ we recover the $c=0$ solution of the previous model.

Now, a phase transition will occur due to changes in the sign of the Hessian of the utility, which is given by
\begin{equation}
    \frac{\partial^2 U}{\partial {x_i}^2}\bigg\rvert_{x_i^*} = - \frac{a_i}{(x_i^*)^2}\Big[1 - J\rho(2\rho - 1) (x_i^*)^{2\rho}\Big].
\end{equation}
Clearly, assuming $J > 0$, we require $\rho > \frac{1}{2}$ regardless of the value of other parameters for a transition to occur. In that case, concentration will indeed occur for $J > J_\infty$ the critical value for the $\beta \to \infty$ solution. For finite $\beta$, the critical value will be given by $J_\mathrm{crit} \geq J_\infty$ just as in our original model.
Numerical simulations (not shown) confirm these predictions. 
\end{document}